\documentclass[
reprint, 
onecolumn,
superscriptaddress, 
amsmath,amssymb, 
aps, 
prb,
]{revtex4-2}

\usepackage[T1]{fontenc}
\usepackage[utf8]{inputenc}
\usepackage{lmodern}
\usepackage[margin=25mm]{geometry}

\usepackage{amsmath,amssymb,amsfonts,bm}

\usepackage{graphicx}
\graphicspath{{images/}}
\usepackage{color}
\usepackage{booktabs}
\usepackage{multirow}

\usepackage[hidelinks]{hyperref}
\hypersetup{
  pdftitle={EQUIVARIANT HAMILTONIAN PREDICTION WITH MANY-BODY MESSAGE PASSING},
  pdfauthor={Chen Qian et al.},
  pdfkeywords={Equivariant graph neural networks, Density functional theory, Kohn-Sham Hamiltonian matrix}
}

\begin{document}


\title{Equivariant Electronic Hamiltonian Prediction with Many-Body Message Passing}

\author{Chen Qian}
\affiliation{Department of Chemistry, University of Warwick, Coventry, CV4 7AL, United Kingdom}
\author{Valdas Vitartas}
\affiliation{Department of Chemistry, University of Warwick, Coventry, CV4 7AL, United Kingdom}
\affiliation{Warwick Centre for Predictive Modelling, School of Engineering, University of Warwick, Coventry, CV4 7AL, United Kingdom}
\author{James R. Kermode}
\affiliation{Warwick Centre for Predictive Modelling, School of Engineering, University of Warwick, Coventry, CV4 7AL, United Kingdom}
\author{Reinhard J. Maurer}\email{reinhard.maurer@univie.ac.at}
\affiliation{Department of Chemistry, University of Warwick, Coventry, CV4 7AL, United Kingdom}
\affiliation{Department of Physics, University of Warwick, Coventry, CV4 7AL, United Kingdom}
\affiliation{University of Vienna, Faculty of Physics, Kolingasse 14-16, Vienna, Austria}


\begin{abstract}
Machine learning surrogate models of Kohn-Sham Density Functional Theory Hamiltonians provide a powerful tool for accelerating the prediction of electronic properties of materials, such as electronic band structures and density of states. For large-scale applications, an ideal model would exhibit high generalization ability and computational efficiency. Here, we introduce the MACE-H graph neural network, which combines high body-order message passing with a node-order expansion to efficiently obtain all relevant $O(3)$ irreducible representations. The model achieves high accuracy and computational efficiency and captures the full local chemical environment features of, currently, up to $f$ orbital matrix interaction blocks. We demonstrate the model's accuracy and transferability on several open materials benchmark datasets of two-dimensional materials and a new dataset for bulk gold, achieving sub-meV prediction errors on matrix elements and high accuracy on eigenvalues across all systems. We further analyze the interplay of high-body-order message passing and locality that makes this model a good candidate for high-throughput material screening.
\end{abstract}

\maketitle

\section*{Introduction}

Machine learning and deep learning have become central tools in computational materials science, offering new ways to accelerate simulations and discover materials~\citep{AI4Science}. A key remaining challenge is how to scale electronic structure calculations to larger time and length scales without sacrificing accuracy. Traditional first-principles methods such as Kohn–Sham density functional theory (KS-DFT) are limited by their steep computational scaling, making them impractical for large or complex systems. Decoupling the accuracy of electronic structure methods from their numerical bottlenecks is essential for enabling predictive simulations across broader domains. Machine learning methods can achieve ab initio accuracy with high computational efficiency, thereby extending applicability to systems with sizes beyond the limits of current approaches.

Machine learning interatomic potentials (MLIPs) have made significant progress in this direction~\citep{mlff_review}, providing efficient and accurate models for atomic interactions~\cite{drautzAtomicClusterExpansion2019,megnet, dpa-1, deepmd, nequip, allegro, MACE, SO3krates}. These models are typically trained on ab initio energy and force data and have demonstrated remarkable transferability and fidelity. These advances have spurred the development of universal potentials, such as MACE-MP-0 \cite{MACE-MP-0} and GNoME \citep{GNoME}. Despite the success of MLIPs in accelerating atomistic simulations, these models fundamentally omit explicit electronic degrees of freedom. As a result, they are limited to predicting structural and thermodynamic properties, and cannot access electronic observables such as band structures, charge transport, or optical responses.

Historically, efforts to make KS-DFT more efficient have involved simplifying the treatment of electronic interactions, either by neglecting certain integrals or by imposing strong approximations on the range or body order of the effective potential. These simplifications underpin semi-empirical methods, which trade physical fidelity for computational speed.~\cite{Seifert07, hourahine_dftb_2020, dral_modern_2024}

\begin{figure*}
    \centering
    \includegraphics[width=\textwidth]{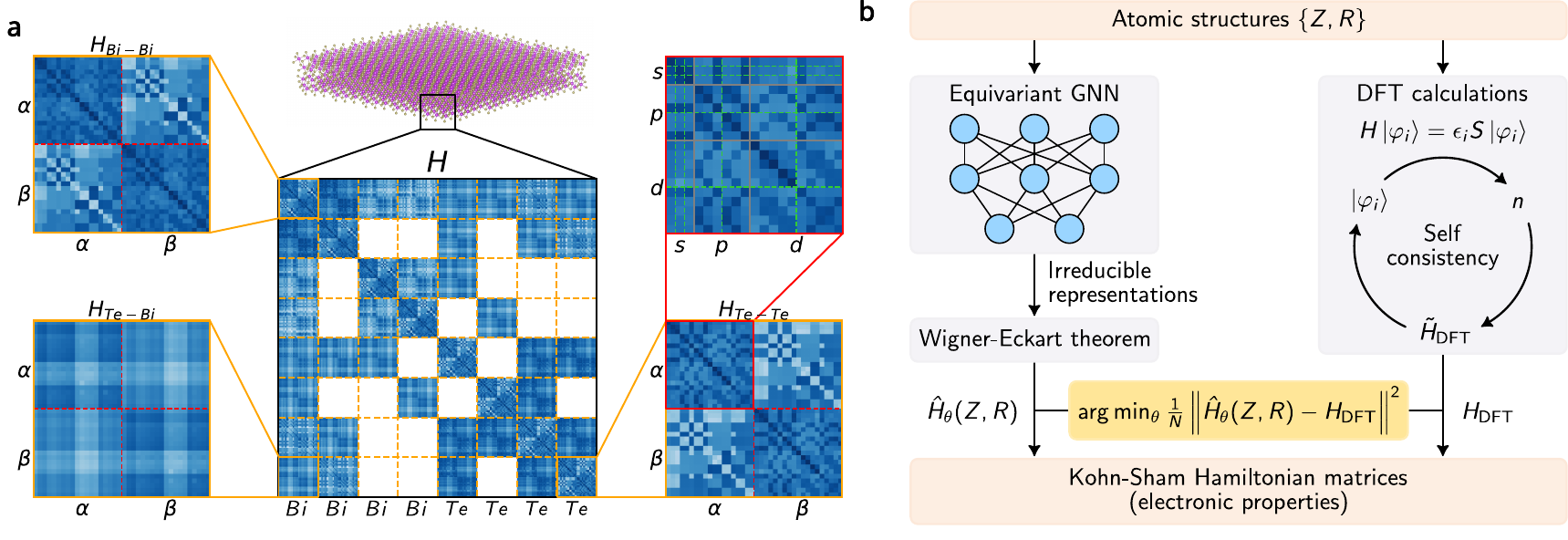}
    \caption{\textbf{Illustration of the Hamiltonian Matrix Prediction}. \textbf{a}. The composition of Hamiltonian matrices for a twisted Bi\textsubscript{2}Te\textsubscript{3} bilayer with spin-orbit coupling, the matrix can be decomposed into atom-pair-wise blocks, which consist of spin-polarized orbital-pair subblocks, \textbf{b}. The comparison of Hamiltonian matrix calculations via neural network and self-consistent field density functional theory. The training of the neural network is supervised by the labels generated from the density functional theory.}
    \label{fig: illustration}
\end{figure*}

An alternative route to increased efficiency has been the development of linear-scaling electronic structure methods. These approaches exploit the locality of atomic orbital basis sets and the sparsity of real-space Hamiltonians. By leveraging the compactness of atom-centered representations, such methods can scale to larger systems while retaining a representation of the electronic structure.~\cite{goedecker_linear_1999,dawson_density_2022} However, their accuracy and transferability remain constrained by the rigidity of their functional forms. These methods rely on the sparsity of the real-space Kohn-Sham (KS) Hamiltonian matrix for their efficiency, as illustrated in Fig.~\ref{fig: illustration}. 

Broadly, ML surrogate models explored to date fall into two categories: tight-binding (TB) methods and direct KS Hamiltonian prediction. Machine learning tight-binding methods predict tight-binding parameters and transform these parameters into matrix elements by Slater-Koster formulas. Due to the scalar nature of Slater-Koster parameters, these models can be easily trained on band structures or electronic density using well-established ML methods like Gaussian process regression (GPR), multilayer perceptrons (MLPs) \citep{kerneltb, Li2018, mcsloy_tbmalt_2023}, and graph neural networks \citep{deeptb}. However, the accuracy of tight-binding methods suffers due to limited body order and parameter approximations, limiting applicability to systems dominated by delocalized interactions.

The second category of approaches uses parametric surrogate models for KS Hamiltonians directly. The first such framework was SchNOrb~\cite{schnorb}, which demonstrated the feasibility of predicting molecular Hamiltonians using deep tensor networks. Further examples including SchNet \citep{schnet}, crystal graph convolutional neural networks (CGCNN) \citep{cgcnn, potnet, matformer, spherenet} have constructed mappings from geometric information to scalar properties with high accuracy. While successful for small molecules, these models typically rely on low-order interactions and are limited in their ability to generalize to extended systems or complex environments.

Our approach in this work builds on recent advances in E(3)-equivariant neural networks, which leverage irreducible representations and tensor product operations to naturally encode rotational symmetries in atomic systems. Equivariant frameworks, such as e3nn and E3x \citep{e3nn, e3x}, can intrinsically handle the symmetry properties of Hamiltonian matrices involving orbitals of arbitrary angular momentum. Compared to invariant GNNs that rely on local coordinate systems—such as SchNOrb~\cite{schnorb} or DeepH \citep{DeepH}, irreps-based equivariant models like PhiSNet \citep{PhiSNET}, QHNet~\citep{QHNet}, SPHNet~\citep{SPHNet}, HamGNN~\citep{HamGNN}, NICE~\cite{nigam_equivariant_2022}, and DeepH-E3  \citep{DeepH-E3} offer enhanced geometric awareness and have demonstrated superior accuracy in electronic structure prediction.

While these models have successfully introduced equivariance into Hamiltonian learning, they typically rely only on two-body messages. In contrast, our model incorporates high-body-order messages into the learning of electronic Hamiltonians, which, for interatomic potentials such as MACE,~\cite{MACE} has shown to reduce the number of required message-passing iterations to achieve similar accuracy as models based on two-body messages. It was also shown that higher body order messages lead to steeper learning curves. High body-order messages have also shown promising results in linear Hamiltonian models based on Atomic Cluster Expansion (ACE) features \citep{drautzAtomicClusterExpansion2019,aceHamiltonians,linearHamiltonians}. Inclusion of high body-order messages captures complex many-body correlations that are essential for modeling extended systems and intricate local environments, such as those found in twisted bilayers and defected bulk materials. This combination of equivariant architecture and high-order interaction modeling promises accurate, transferable predictions of electronic structure across diverse materials classes without having to introduce many more internal parameters.

Specifically, we introduce the MACE-H model, which incorporates the efficient many-body order message-passing scheme from MACE~\cite{MACE} into a graph neural network representation of KS DFT Hamiltonians. Taking advantage of the increased body order in the message passing phase, the model achieves high prediction accuracy and computational and data efficiency. MACE-H achieves high expressiveness and is compatible with spin-orbit coupling and multi-element systems. The performance of MACE-H was benchmarked across various datasets against the two-body equivariant message-passing GNN DeepH-E3.
The datasets include 2D shifted and twisted bilayers and 3D bulk materials using metrics based on matrix elements, band structures, and density of states, such as the eigenvalue error and electronic entropy error. The latter tests show that high accuracy in matrix elements persists in downstream property predictions based on the eigenvalues. Further tests reveal the interplay between the many-body expansion of messages and the emphasis of the model towards messages closer or further away from the central atom or atom pair. In addition to the many-body message-passing scheme, we also employed a shift and scale operation that increases accuracy, stability and accelerates convergence. Finally, we found that the model prediction error can be estimated by the Hermiticity of the model output, which is proposed as a possible metric during active learning. While the model is designed for the prediction of KS DFT Hamiltonians, it can be more generally employed for the prediction of quantum operators in local basis representation, which will benefit high-throughput materials discovery, machine-learning augmented electronic structure theory, and coupled electron-nuclear dynamics simulations.

\section*{Results}

\subsection*{The MACE-H model}
\begin{figure*}
    \centering
    \includegraphics[width=0.95\textwidth]{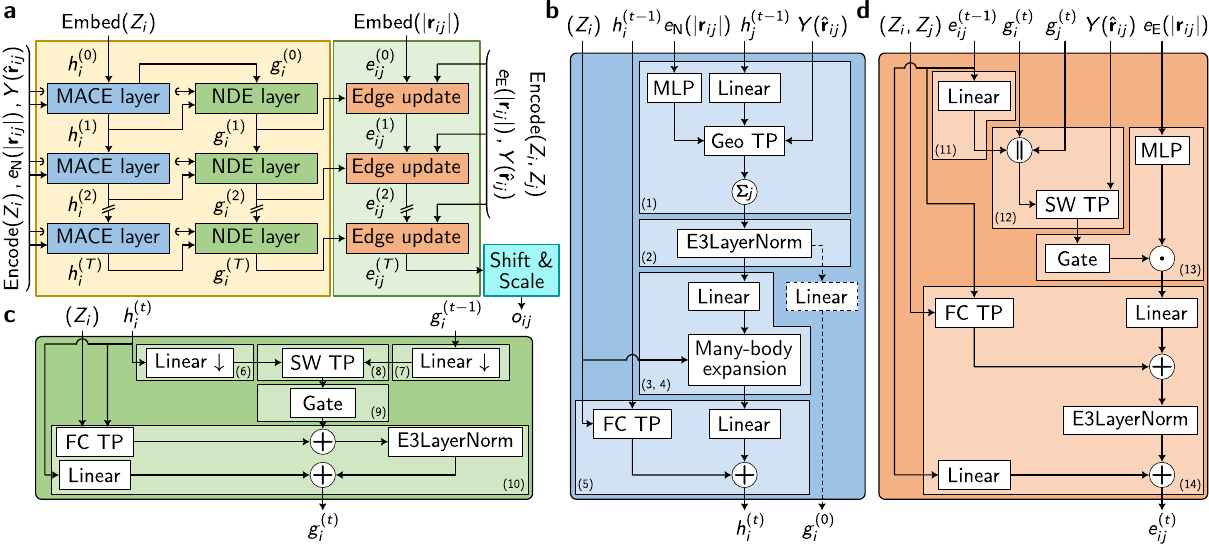}
    \caption{\textbf{Graph Neural Network Architecture of MACE-H.} \textbf{a.} The overall workflow of module blocks. The node-wise message passing operations (as denoted in the yellow rectangle) consist of $L$ message neural network layers. $\mathrm{Encode(}Z_i\mathrm{)}$ and $\mathrm{Encode(}Z_i, Z_j\mathrm{)}$
    are one-hot-encoding of the element $Z$ of atom $i$ and the element pair $Z_i-Z_j$ between atom $i$ and $j$, respectively. The initial node feature in MACE block and edge feature in the edge-update block in the first layer are the element embedding $\mathrm{Embed}\left(Z_{i}\right)$ and distance embedding $\mathrm{Embed}\left(\left|\bm{r}_{ij}\right|\right)$. $e_{\mathrm{N}}\left(\left|\bm{r}_{ij}\right|\right)$ and $e_{\mathrm{E}}\left(\left|\bm{r}_{ij}\right|\right)$ stand for node-update and edge-update radial basis sets, respectively. \textbf{b.} The MACE layer is used to aggregate the atom-level chemical environment feature representation, which encodes atom embedding, edge length, and orientation information to higher-body features. \textbf{c.} The node degree expansion (NDE) layer elevates the node feature degree in order to be compatible with the edge-wise irreducible representations (irreps) corresponding to the Hamiltonian subblocks. \textbf{d.} The edge update block converts the node-wise features and geometry information into the edge-wise features corresponding to the matrix block. This is identical to the edge update block in DeepH-E3 \cite{DeepH-E3}. All abbreviations are defined in the main text.}
    \label{fig: workflow}
\end{figure*}

The MACE-H graph neural network includes three different modules: 1) the node-wise (atom-wise) MACE message passing, 2) the node degree expansion, and 3) the edge-wise message updating. It will be shown below that the many-body expansion increases data efficiency and prediction accuracy while maintaining high expressiveness in the message passing phase. The overall workflow is illustrated in Fig.~\ref{fig: workflow}. The specific implementation of the three modules is explained in the following sections.

The node-wise message passing, used to obtain the atom-wise feature representation, is implemented by the MACE block. Similar to the implementation of the MACE interatomic potential~\citep{MACE}, the message-passing phase of atom $i$ produces an intermediate atom feature $A^{(t)}_{i}$ with 2-body interaction. For this, a message function, $\tilde{A}_{i,kl_{3}m_{3}}^{(t)}$, is produced by aggregation of the atomic neighborhood $\mathcal{N}(i)$ at layer $t$:
\begin{align} \label{eqn: Geo TP}
\tilde{A}_{i,kl_{3}m_{3}}^{(t)}=\sum_{j\in{\mathcal{N}(i)}}\sum_{l_{1}m_{1},l_{2}m_{2}}^{l_{3}m_{3}}R_{kl_{1}l_{2}l_{3}}^{(t)}(|\bm{r}_{ji}|)C_{l_{1}m_{1},l_{2}m_{2}}^{l_{3}m_{3}}Y_{l_{1}}^{m_{1}}(\hat{\bm{r}}_{ji})\sum_{\tilde{k}}W_{k\tilde{k}l_{2}}^{(t)}h_{j,\tilde{k}l_{2}m_{2}}^{(t-1)}\,,
\end{align}
where $R_{kl_{1}l_{2}l_{3}}^{(t)}$ is a learnable weight of each Clebsch-Gordan (CG) tensor product path, which is obtained by a multilayer perceptron (MLP) taking the Bessel or Gaussian radial basis embedding of the interatomic distance $|\bm{r}_{ji}|$ as input, $C_{l_{1}m_{1},l_{2}m_{2}}^{l_{3}m_{3}}$ is the CG coefficient to retain the equivariance of the CG tensor products, $Y_{l_{1}}^{m_{1}}(\hat{\bm{r}}_{ji})$ is a spherical harmonic function of the interatomic vector $\hat{\bm{r}}_{ji}$, where the hat indicates a unit vector. The learnable weights $W_{k\tilde{k}l_{2}}^{(t)}$ linearly transform hidden states $h_{j,\tilde{k}l_{2}m_{2}}^{(t-1)}$ from a previous layer. Eq.~\ref{eqn: Geo TP} is implemented via a geometric tensor product (Geo TP) following the same formalism as NequIP \citep{nequip}. The Geo TP between irreps features and spherical harmonics is used to incorporate geometric information because it has fewer tensor product paths than the fully connected tensor product (FC TP), thereby avoiding overparameterization and computational overhead. This implementation is equivalent to the depth-wise tensor product in Equiformer \citep{Equiformer, Equiformerv2}. As a special case, in the first layer, Eq.~\ref{eqn: Geo TP} is simplified with $h_{j,\tilde{k}l_{2}m_{2}}^{(0)}$ being a one-hot embedding of the element $Z_j$.

In interatomic potentials, only the energies as scalar outputs ($l=0$) are the final prediction results. For Hamiltonian predictions, we require irreps with higher degrees ($l>0$) to construct Hamiltonian subblocks with high numerical stability. To facilitate model convergence and numerical stability of the gradient, MACE-H directly adopts summation as the aggregation scheme without division by the average node degree (as used in MACE), and also employs the E(3) layer normalization operation in DeepH-E3 \citep{DeepH-E3}:
\begin{align}
A_{i,kl_{3}m_{3}}^{(t)}=\text{E3LayerNorm}\left(\tilde{A}_{i,kl_{3}m_{3}}^{(t)}\right) =W_{kl_{3}}^{(t)}\frac{\tilde{A}_{i,kl_{3}m_{3}}^{(t)}-\mu_{kl_3}^{(t)}}{\sigma_{l_3}^{(t)}+\varepsilon}+b_{kl_{3}}^{(t)}\,,
\end{align}
where $W_{kl_3}$ and $b_{kl_3}$ are learnable affine parameters, but $b_{kl_3}$ is only used for scalars. The mean values $\mu_{kl_3}^{(t)}=\frac{1}{Nn}\sum_{i=1}^{N}\sum_{k=1}^{n}\tilde{A}_{i,kl_{3}m_{3}}^{(t)}$ are calculated separately for each irrep, $l_3$, by averaging across the $N$ nodes in the same graph and $n$ channels. The standard deviation $\sigma_{l_3}^{(t)}=\sqrt{\frac{1}{Nn}\sum_{i=1}^{N}\sum_{k=1}^{n}\sum_{m_{3}=-l_3}^{l_3}{\left\|\tilde{A}_{i,kl_{3}m_{3}}^{(t)}-\mu_{kl_3}^{(t)}\right\|}^2}$ is only active for scalars in MACE-H. $\varepsilon$ is a small number in the denominator to guarantee numerical stability. 

Higher-order $(\nu+1)$-body information is introduced with the many-body expansion operation on the node-wise messages, where messages, labelled with $\eta_{\nu}=(l_1, \dots, l_\nu)$, with a given correlation order $\nu$ contain irreps up to $l_{\nu}$:
\begin{align}\label{eqn: mbe}
B_{i,\eta_{\nu}kLM}^{(t)}&=\sum_{\textit{\textbf{lm}}}\mathcal{C}_{\eta_{\nu},\textit{\textbf{lm}}}^{LM}\prod_{\xi=1}^{\nu}\sum_{\tilde{k}}W_{k\tilde{k}l_\xi}^{(t)}A_{i,\tilde{k}l_{\xi}m_{\xi}}^{(t)}, \quad
\textit{\textbf{lm}}=(l_1m_1, \dots, l_{\nu}m_{\nu})\,
\end{align}
\begin{equation}\label{eqn: mbe_lin}
m_{i,kLM}^{(t)}=\sum_{\nu}\sum_{\eta_\nu}W_{Z_{i}kL,\eta_{\nu}}^{(t)}B_{i,\eta_{\nu}kLM}^{(t)}\,,
\end{equation}
where the $W_{k\tilde{k}l_\xi}^{(t)}$ is the learnable weight for the linear transformation, the generalized CG coefficients are calculated based on the standard CG coefficients and the related indices to symmetrize the many-body features $B_{i,\eta_{\nu}kLM}^{(t)}$. Here, we employed the loop tensor contraction algorithm as in the MACE potential for the calculation of Eq.~\ref{eqn: mbe} and ~\ref{eqn: mbe_lin}, the more detailed implementation  and the calculation of generalized CG coefficients can be found in Ref.~\citep{MACE}. $W_{Z_{i}kL,\eta_{\nu}}^{(t)}$ is the learnable element-dependent weight to perform the linear combination of $B_{i,\eta_{\nu}kLM}^{(t)}$ and obtain the feature $m_{i,kLM}^{(t)}$. 

The hidden state of each layer $h_{i,kLM}^{(t)}$ is obtained by the residual connection of the layer input, i.e., the last layer hidden state $h_{i,kLM}^{(t-1)}$, and corresponding linear transformations over the channels:
\begin{equation}
h_{i,kLM}^{(t)}=\sum_{\tilde{k}}W_{Lk\tilde{k}}^{(t)}m^{(t)}_{i,\tilde{k}LM}+\sum_{\tilde{k}}W_{Z_{i}Lk\tilde{k}}^{(t)}h_{i,\tilde{k}LM}^{(t-1)}\,.
\end{equation}

Next, we discuss the node degree expansion block. One significant characteristic of the KS Hamiltonian matrix prediction task is the presence of irreps with high angular momentum. For instance, a $f-f$ matrix subblock with spin-orbit coupling (SOC) will require edge irreps with maximum degree $l_{\mathrm{max}}=9$ according to the Wigner-Eckart theorem, which is intractable for the CG tensor product in the implementation of the message function and many-body expansion. Accordingly, the largest angular momentum $l_{\mathrm{max}}$ of the node irreps should also be larger than or equal to the summation of the angular momenta of the two orbitals in an orbital pair, since the node irreps will not only be processed to obtain the off-diagonal blocks of different atom pairs but will also be used to calculate the diagonal blocks of the atom itself. Moreover, as the generalized CG coefficients are pre-calculated with high sparsity, the size increases exponentially with the increase in angular momentum, which requires excessive floating-point operations and GPU memory. As a result, the many-body expansion operation is not straightforwardly applicable to KS Hamiltonian prediction without adaptation. To this end, we introduce the node degree expansion block to increase the degree of features with affordable node-wise operations, which bridge the angular momentum mismatch between edge-wise high-degree irreps and node-wise hidden states created by the many-body expansion.

The core implementation is the tensor product between two irreps with one being the hidden states $h_{i,kLM}^{(t)}$ from the same layer MACE block and the other being the $\tilde{A}_{i,kl_{3}m_{3}}^{(1)}$ for the first layer and the last node degree expansion layer outputs $g_{i,klm}^{(t-1)}$ for number of layers $t\geq2$. To obtain the higher degree intermediate feature $E_{i,kL_{3}M_{3}}^{(t)}$,  a separate-weight tensor product (SW TP) acts on the two inputs that are linearly transformed to achieve downsampling in the number of channels: 

\begin{align}
\bar{h}_{i,k_{1}L_{1}M_{1}}^{(t)} = \sum_{\tilde{k}_{1}} \bar{W}_{L_{1}k_{1}\tilde{k}_{1}}^{(t)} h_{i,\tilde{k}_{1}L_{1}M_{1}}^{(t)}\,, 
\end{align}

\begin{align}
\bar{g}_{i,k_{2}L_{2}M_{2}}^{(t-1)} = \sum_{\tilde{k}_{2}} \bar{W}_{L_{2}k_{2}\tilde{k}_{2}}^{(t)} g_{i,\tilde{k}_{2}L_{2}M_{2}}^{(t-1)}\,,
\end{align}

\begin{align}
E_{i,kL_{3}M_{3}}^{(t)} = \sum_{L_{1}M_{1},L_{2}M_{2}}^{L_{3}M_{3}}\,\sum_{k_{1},k_{2}} 
W_{L_{1},kk_{1}}^{(t)}W_{L_{2},kk_{2}}^{(t)}
C_{L_{1}M_{1},L_{2}M_{2}}^{L_{3}M_{3}}
\bar{h}_{i,k_{1}L_{1}M_{1}}^{(t)} \bar{g}_{i,k_{2}L_{2}M_{2}}^{(t-1)}\,,
\end{align}
where $\bar{W}_{L_{1}k_{1}\tilde{k}_{1}}^{(t)}, \bar{W}_{L_{2}k_{2}\tilde{k}_{2}}^{(t)}$ and $W_{L_{1},kk_{1}}^{(t)}, W_{L_{2},kk_{2}}^{(t)}$ correspond to downsampling and tensor product weights, respectively. Compared to the FC TP, the SW TP can effectively avoid overparameterization of the TP, especially when the input features have more channels. 

For higher nonlinearity, we adopted the gate activation for $E_{i,kL_{3}M_{3}}^{(t)}$:
\begin{align} \label{eqn: GateActivation}
F_{i}^{(t)} =\mathrm{Gate}(E_{i}^{(t)}) =\left(\underset{L_{1}=0}{\oplus}\phi\left(E_{\dot{k}L_{1}}^{(t)}\right)\right)\oplus\left(\underset{L_{2}=0,\,L_{3}\geq1}{\oplus}\varphi\left(E_{\ddot{k}L_{2}}^{(t)}\right)E_{\hat{k}L_{3}}^{(t)}\right)\,,
\end{align}
where $\oplus$ denotes the direct sum of irreps, $E_{\dot{k}L_{1}}^{(t)}$ corresponds to the scalar components of the outputs, $E_{\ddot{k}L_{2}}^{(t)}$ is the gating scalar, $E_{\hat{k}L_{3}}^{(t)}$ is the nonscalar component of the irreps, $\phi$ and $\varphi$ are activation functions. $\phi$ is the SiLU and tanh activation function for even and odd parity scalars, respectively. For $\varphi$, the sigmoid and tanh activation functions are used for gating scalars with even and odd parity, respectively.

The output $g_{i,kLM}^{(t)}$ is obtained through linear transformation and residual connection to the hidden states $h_{i,\tilde{k}LM}^{(t)}$:
\begin{align}
g_{i,kLM}^{(t)} =\mathrm{E3LayerNorm}\left(
F_{i,kLM}^{(t)}+\sum_{\tilde{k}}W_{Z_{i}Lk\tilde{k}}^{(t)}h_{i,\tilde{k}LM}^{(t)}\right) +\sum_{\tilde{k}}W_{Lk\tilde{k}}^{(t)}h_{i,\tilde{k}LM}^{(t)}\,.
\end{align}
In this equation, NDE inputs $h_{i,\tilde{k}l_3m_3}^{(t)}$ with lower degree are padded with zeros to a higher degree $LM$
as they have lower maximum angular momentum than the $L$ and $M$ created through the node degree expansion.

In MACE-H, we directly implement the edge update block previously proposed as part of the DeepH-E3 model \citep{DeepH-E3}. The edge-wise feature $e_{ij}^{(t)}$ in each layer is updated by the corresponding node features, edge vector, and last layer output. Intermediate edge features, $S_{ij}^{(t)}$, are created by linear transformation of the last layer edge update output $e_{ij}^{(t-1)}$. These are concatenated with the related node features $g_{i}^{(t)}$ and $g_{j}^{(t)}$ and updated features $U_{ij,kl_{3}m_{3}}^{(t)}$ are calculated with a separate-weight CG tensor product with corresponding spherical harmonics $Y(\hat{\bm{r}}_{ji})$:
\begin{equation}
S_{ij,klm}^{(t)}=\sum_{\tilde{k}}W_{lk\tilde{k}}^{(t)}e_{ij,\tilde{k}lm}^{(t-1)}\,,
\end{equation}
\begin{align}
U_{ij,kl_{3}m_{3}}^{(t)}=
\sum_{l_{1}m_{1},l_{2}m_{2}}^{l_{3}m_{3}}
\sum_{\tilde{k}}
W_{l_{1},k}^{(t)}W_{l_{2},k\tilde{k}}^{(t)}
C_{l_{1}m_{1},l_{2}m_{2}}^{l_{3}m_{3}}
Y_{l_{1}}^{m_{1}}(\hat{\bm{r}}_{ji}) 
\left(\left.g_{i}^{(t)}\right\Vert \left.g_{j}^{(t)}\right\Vert S_{ij}^{(t)}\right)_{\tilde{k}l_{2}m_{2}}\,,
\end{align}
where the $e_{ij,klm}^{(0)}$ as input of the first layer is initialized by the edge distance embedding. A gate activation similar to Eq.~\ref{eqn: GateActivation} is also employed on the edge information $U_{ij}^{(t)}$ to increase model nonlinearity. Then, an element-wise multiplication is performed between each irrep and length-dependent weight to incorporate the distance information, while the weight is parameterized as the MLP of Gaussian radial basis function (as in Eq.~\ref{eqn: Geo TP}):
\begin{equation}
P_{ij,klm}^{(t)}=\mathrm{Gate}\left(U_{ij}^{(t)}\right)_{klm}W_{kl}^{(t)}(|\bm{r}_{ji}|)\,.
\end{equation}
For the final output of the block, linear transformation, residual connection and layer normalization are employed:  
\begin{align}  &e_{ij,klm}^{(t)}= \mathrm{E3LayerNorm}\left(\sum_{\tilde{k}}W_{lk\tilde{k}}^{(t)}P_{ij,\tilde{k}lm}^{(t)}+\sum_{\tilde{k}}W_{Z_{ij}lk\tilde{k}}^{(t)}e_{ij,\tilde{k}lm}^{(t-1)}\right) +\sum_{\tilde{k}}\bar{W}_{lk\tilde{k}}^{(t)}e_{ij,\tilde{k}lm}^{(t-1)}\,,
\end{align}
where the $W_{Z_{ij}}$ are element-pair-wise learnable weights.

Upon the model output of the corresponding irreps $o_{ij,kl_{3}m_{3}}$ in the direct sum form, each orbital-pair-resolved subblock in tensorial form can be converted to the Cartesian Hamiltonian representation according to the Wigner-Eckart theorem using the tensor product expansion. For occasions without SOC, the conversion is:

\begin{equation}     
H_{ij, l_{1}m_{1}l_{2}m_{2}}=\sum_{l_{3} = |l_{1} - l_{2}|}^{l_{1} + l_{2}}\sum_{m_{3}=-l_{3}}^{l_{3}} C_{l_{3}m_{3}}^{l_{1}m_{1},l_{2}m_{2}}o_{ij,kl_{3}m_{3}}\,,
\end{equation}
where $C$ is also the CG coefficient, this step can be regarded as the inverse operation of the tensor product contraction. The matrix form output represents the component with different group symmetries of the corresponding subblock. Notably, the number of channels in each segment of the edge-wise output irreps is predetermined according to the Wigner-Eckart theorem, such that all of them are used to construct the Hamiltonian matrix. Subblock matrices are then obtained as the summation of the tensor product form of the relevant irreps. Considering that in multi-element systems, different elements may have different orbital angular momentum of the basis set which leads to different types of orbital pair subblocks in different element pair blocks, the data pipeline loads edgewise block label in such a way that the edgewise matrix blocks will be split into different orbital pair-resolved subblocks, with the subblocks flattened into different segments of a 1D feature and aligned according to the angular momenta of the orbital pairs. The missing orbital pair components in the label of certain element pairs will be padded to zero and another mask tensor of the same shape will be employed to indicate which components are padded and should be excluded during loss function calculations. This will facilitate efficient conversion between irreps in the format of direct sum and direct product. For the SOC scenario, the matrix elements are complex values and there are blocks between different spin channels. The model structure for SOC Hamiltonian treats the irreps feature in the same way as the noSOC  situation, except that it incorporates a different conversion operation from the model outputs to the matrix form, while different halves of the real value model output correspond to the real and imaginary components of the complex value. This tensor product expansion operation is taken from DeepH-E3, and the implementation details of the SOC cases can be found in Ref.~\citep{DeepH-E3}. The sparse overall Hamiltonian matrices are assembled from the subblocks for downstream calculations.

\subsection*{Model Validation and Benchmark}

\begin{figure*}
    \centering
    \includegraphics[width=\textwidth]{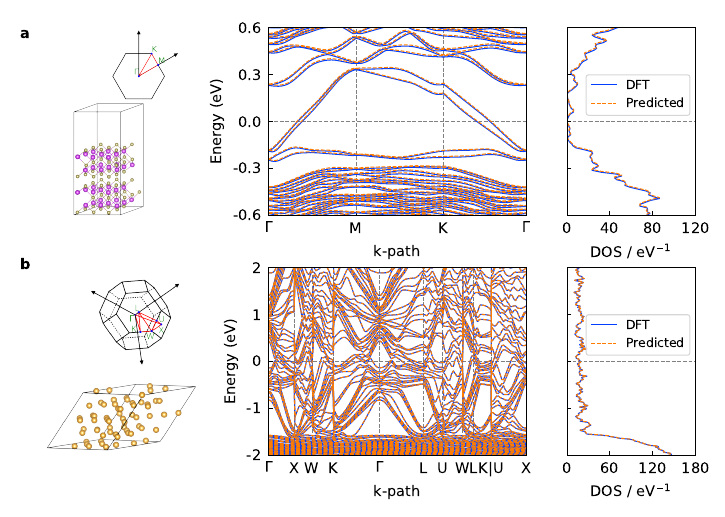}
    \caption{\textbf{Demonstration of MACE-H for Materials Electronic Structure Prediction.} The predicted band structure along the high-symmetry k-path and the density of states in comparison with DFT results for a two-dimensional bilayer of Bi\textsubscript{2}Te\textsubscript{3} (\textbf{a}) and face-centered-cubic bulk Au (\textbf{b}).}
    \label{fig: demonstration}
\end{figure*}

The performance of the MACE-H model was assessed by training the model separately on DFT data for a two-dimensional bilayer of Bi\textsubscript{2}Te\textsubscript{3} and face-centered-cubic bulk Au. Details on the training data are provided in the Methods section. Both systems represent good benchmarks as Bi\textsubscript{2}Te\textsubscript{3} features strong SOC with high geometrical complexity, and Au KS Hamiltonians calculated with all-electron DFT calculations require the model to handle many Hamiltonian blocks with high accuracy and numerical stability. To compare pseudopotential-based and all-electron predictions on equal footing and to minimise the overhead of predicting matrix blocks associated with core states, we employ a core projection on the bulk gold data (Methods sections \nameref{sec:data_generation} and \nameref{sec:core_projection}). 

MACE-H achieves an overall mean absolute error (MAE) of 0.278 meV with shift-and-scale operation for the Hamiltonian matrix elements of a hold-out test set for the shifted Bi\textsubscript{2}Te\textsubscript{3} dataset with SOC. Detailed orbital pair-resolved error matrices are shown in Supplementary Fig. 1. While direct MAEs on Hamiltonian matrix elements have previously been commonly used to assess model performance in literature, we note they are not instructive in assessing the ability of the model to faithfully reproduce DFT-level band structures or densities-of-states (DOS) of materials. To quantify the performance in electronic property prediction, the eigenvalue error (EE) and electronic entropy error (EEE) metrics have been defined and employed (Methods section \nameref{sec:metrics}). The corresponding values of EE and EEE for the hold-out test set  Bi\textsubscript{2}Te\textsubscript{3} configuration shown in Fig.~\ref{fig: demonstration}a at a temperature of 1000 K are 14.6 meV and $7.53\times 10^{-8}$ meV$\cdot$K\textsuperscript{-1}$\cdot$\text{\AA}\textsuperscript{-3}, respectively.

For the Au dataset, MACE-H with shift-and-scale operation reached an MAE of 0.269 meV for the Hamiltonian matrix elements in the test set. The corresponding EE and EEE errors for the test Au configuration in Fig.~\ref{fig: demonstration}b at 1000 K are 2.5 meV and $2.19\times 10^{-7}$ meV$\cdot$K\textsuperscript{-1}$\cdot$\text{\AA}\textsuperscript{-3}. For both systems, these are errors that produce bandstructures and DOS that are visually indistinguishable from the DFT reference calculations (see Fig.~\ref{fig: demonstration}).

\begin{figure*}
    \centering
    \includegraphics[width=\textwidth]{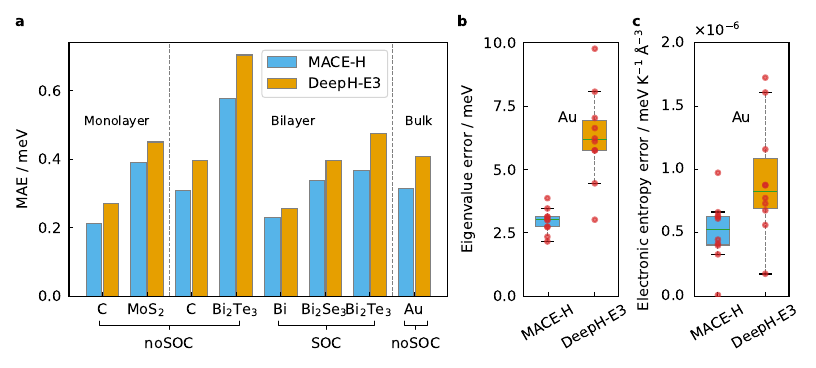}
    \caption{\textbf{Model Performance Assessment of MACE-H Compared to DeepH-E3.} \textbf{a.} The test set mean absolute error of matrix elements for different datasets of 2D monolayers, shifted 2D bilayers, and bulk systems. The 2D material datasets are generated by the OpenMX package, the Au all-electron dataset is generated with FHI-aims. \textbf{b.} The eigenvalue error and \textbf{c.} electronic entropy error for Au datasets. The eigenvalue error and electronic entropy error are defined in Section \nameref{sec:metrics}. The hyperparameters are listed in Supplementary Table 3. The correlation order is $\nu=3$.}
    \label{fig: benchmark}.
\end{figure*}

To directly compare MACE-H against the message passing model DeepH-E3, which has a similar architecture but only employs two-body messages, MACE-H and DeepH-E3 were trained on a previously published 2D materials dataset for materials with SOC and without SOC (noSOC)~\citep{DeepH-E3, DeepH}. This dataset includes DFT Kohn-Sham Hamiltonians calculated with OpenMX \citep{OpenMx1, OpenMx2} for noSOC monolayer and bilayer graphene, noSOC monolayer MoS\textsubscript{2}, SOC bilayer bismuthene, SOC bilayer Bi\textsubscript{2}Se\textsubscript{3}, and SOC/noSOC bilayer Bi\textsubscript{2}Te\textsubscript{3}. The training data was generated by random displacements and interlayer shifting, with details explained in Methods section \nameref{sec:data_generation}. Moreover, we employed the same hyperparameter settings of the edge-update blocks for both the MACE-H and DeepH-E3 to illustrate the effectiveness of the higher-body messages of MACE-H, and also adopted similar settings for the message-passing process of MACE-H (with correlation order $\nu=3$) and DeepH-E3, as hyperparameters available in Supplementary Table 3 and Section \nameref{sec: data availability}.

On randomly split held-out test sets, MACE-H shows consistently lower MAE of matrix element prediction compared to DeepH-E3 (Fig.~\ref{fig: benchmark}a, Supplementary Tables 4 and 5). This is the case for both noSOC and SOC data. As shown for the   Bi\textsubscript{2}Te\textsubscript{3} and bulk gold dataset in the Supplementary Material (Supplementary Table 6, Supplementary Figs. 2, and 3), this increase in accuracy holds independently of the correlation order $\nu$, the choice of basis function, and the number of layers. We attribute the improvement to the NDE block. Apart from higher nonlinearity, the NDE block relies on the tensor product between two node-level features to enrich the irreps with higher degrees or different parities, which intrinsically incorporates the 3-body messages regardless of the correlation order $\nu$ in MACE blocks. Consequently, MACE-H can, in fact, still achieve higher body messages even if $\nu=1$, and outperforms DeepH-E3 for the shifted bilayer (see Supplementary Tables 6 and 7 for Bi\textsubscript{2}Te\textsubscript{3} and Bi\textsubscript{2}Se\textsubscript{3}). In addition to benchmarking on the 2D materials and the bulk gold dataset, a detailed evaluation on molecular datasets with comparisons to previous models is discussed in Supplementary Note~3. In essence, MACE-H performs as well or better than models such as PhiSNet and QHNet if provided with comparable training data.

Improved accuracy for MACE-H compared to DeepH-E3 holds for the mono- and bilayer data and the bulk gold data. For the bulk gold system, the pristine MACE-H also achieved a lower MAE on the matrix elements of 0.316 meV, compared to 0.409 meV of DeepH-E3 (Supplementary Table 8). Further improvement of MACE-H by shift-and-scale the output irreps will be discussed in Section \nameref{sc: shift and scale}. To make a fair comparison with DeepH-E3, all figures except Fig.~\ref{fig: demonstration} use MACE-H models trained without the shift-and-scale operation. As assessments for downstream electronic property prediction for the bulk gold system, the average EE and EEE by MACE-H are 2.96 meV and $5.11\times 10^{-7}$ meV$\cdot$K\textsuperscript{-1}$\cdot$\text{\AA}\textsuperscript{-3} compared to 6.28 meV and $9.17\times 10^{-7}$ meV$\cdot$K\textsuperscript{-1}$\cdot$\text{\AA}\textsuperscript{-3} of those by DeepH-E3  for the configurations in Fig. \ref{fig: benchmark}b and \ref{fig: benchmark}c. Since we adopted the same edge update block for MACE-H as was used in DeepH-E3, we attribute the improved expressiveness to the many-body expansion in the message-passing phase of MACE-H, while DeepH-E3 only features 2-body messages.

\subsection*{Data Efficiency and Computational Cost}

\begin{figure}
    \centering
    \includegraphics[width=0.8\textwidth,keepaspectratio]{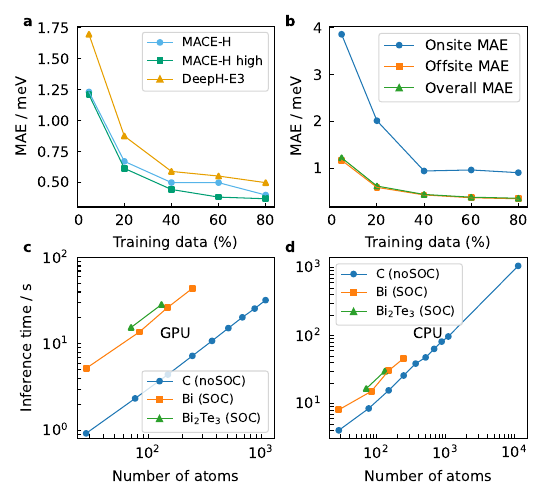}
    \caption{\textbf{Data- and Computational Efficiency of MACE-H.} \textbf{a.} The matrix element MAE as a function of training data size of shifted  Bi\textsubscript{2}Te\textsubscript{3} for MACE-H and DeepH-E3. Here, DeepH-E3 uses spherical harmonics with $l$ up to 4, MACE-H uses the same settings in the edge-update block and node-wise blocks with spherical harmonics with $l$ up to 4, and hidden states with azimuthal number up to 2. The MACE-H with notation "high" uses spherical harmonics $l$ up to 5 and hidden state azimuthal numbers up to 5. The correlation order $\nu=3$ for both MACE-H models. \textbf{b.} The onsite and offsite-resolved matrix element MAE as a function of  Bi\textsubscript{2}Te\textsubscript{3} training data size, which consists of a total of 256 configurations with 90 atoms in each. Furthermore, shown are the single batch inference time of MACE-H on a single GPU (see \textbf{c}) and 64 CPU cores (see \textbf{d}) for graphene bilayers without spin-orbital coupling and Bismuthene and Bi\textsubscript{2}Te\textsubscript{3} bilayers with spin-orbital coupling.}
    \label{fig: time-to-solution}
\end{figure}


To evaluate the influence of the many-body expansion on data efficiency, we compared the MACE-H and DeepH-E3 models with the same setting of azimuthal numbers in irreps. Note that a MACE-H setting with $\nu=1$ doesn't necessarily correspond to a pure 2-body message due to the NDE as explained in the last section. Fig.~\ref{fig: time-to-solution}a shows that with the model depth equal to 3 layers ($T=3$), MACE-H yields lower MAEs than DeepH-E3, and further improvement can be achieved by increasing the azimuthal number of irreps in the hidden states. Here, DeepH-E3 uses spherical harmonics with $l$ up to 4, MACE-H uses the same settings in the edge-update block and node-wise blocks with spherical harmonics with $l$ up to 4, and hidden states with azimuthal number up to 2. The higher setting of MACE-H ("MACE-H high" in Fig.~\ref{fig: time-to-solution}) uses spherical harmonics $l$ up to 5 and hidden state azimuthal numbers up to 5. The correlation order $\nu$  equals 3 for both MACE-H models. With decreasing training data size, MACE-H retains a lower MAE than DeepH-E3 due to the higher expressiveness of the many-body expansion. For example, the MACE-H model with higher azimuthal numbers of irreps and trained on 20\% of the data has an equivalent accuracy as DeepH-E3 trained with 40\% of the data. With the higher-body-order message involved, MACE-H can accurately describe the chemical environment using fewer layers. Supplementary Fig. 4 considers different model depths. The 2-layer MACE-H provides higher accuracy and data efficiency than the 3-layer DeepH-E3. The difference between the same model with different depths decreases with the training data size. Among these, the MACE-H model with higher azimuthal numbers showed a marginal difference between the model depths of 2 and 3 across various data sizes. Furthermore, the 2-layer MACE-H with larger azimuthal numbers outperforms the 3-layer MACE-H with lower ones in terms of data efficiency and accuracy, indicating that increasing the azimuthal number of the irreps is more effective than increasing the model depth. The on-site blocks are the blocks between the same atoms in the central unit cell, while the rest are off-site blocks. On-site blocks converge faster with the number of training data provided than the off-site blocks (Fig.~\ref{fig: time-to-solution}b), but they converge to larger MAE values. This is likely due to the large magnitude of Hamiltonian matrix elements on these blocks.

Fig. \ref{fig: time-to-solution}c and \ref{fig: time-to-solution}d show the inference time for different 2D material datasets using a single GPU and 64 CPU cores, respectively. The inference time of MACE-H is linear with the number of atoms in the system, manifesting the scalability to larger systems. For example, the inference time for a graphene bilayer with 1084 atoms takes 32.1 seconds on an Nvidia A100 GPU. Further acceleration on the GPU can be achieved by batched samples. MACE-H GPU inference only negligibly differs from DeepH-E3, as they share the same edge-update block, which accounts for most of the computational overhead in the model (Supplementary Fig. 5). Meanwhile, the additional many-body expansion provides no significant overhead. The inference time on a CPU is slower than on a GPU, but it is still efficient compared to outright DFT calculations. Due to current memory limitations on available GPU hardware, we perform the Hamiltonian matrix prediction for magic-angle graphene on a CPU. This takes ca. 18 minutes for a system of  11,164 atoms. The computational cost and scaling associated with Hamiltonian diagonalisation remains unaffected by the machine learning prediction and, assuming optimal integration with existing libraries, can only be as efficient as the current state-of-the-art diagonalisation algorithms.

\subsection*{The Effect of Many-Body Expansion}

\begin{figure*}
    \centering
    \includegraphics[width=\textwidth]{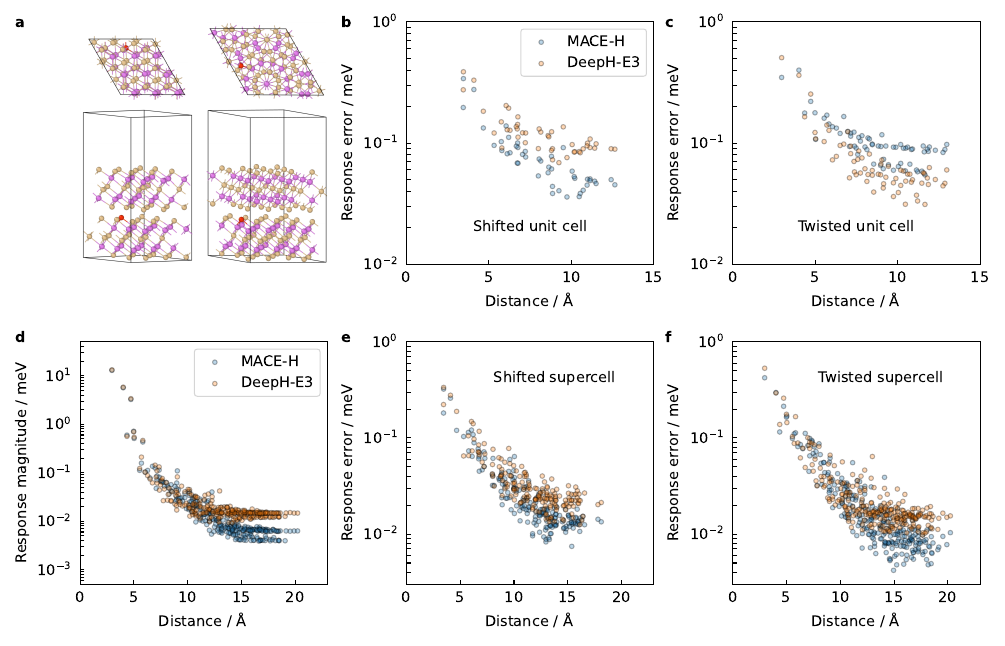}
    \caption{\textbf{Analysis of Atom Perturbation Response of MACE-H and DeepH-E3.} \textbf{a.} The configuration of shifted and twisted Bi\textsubscript{2}Te\textsubscript{3} bilayer unit cells with the red atom being the perturbed Te atom. \textbf{b.} The response error of onsite matrix blocks w.r.t. the distance from the perturbed atom for a shifted bilayer unit cell. \textbf{c.} The onsite response error for a twisted bilayer unit cell. \textbf{d.} The decay rate comparison for the 2$\times$2$\times$1 twisted supercell bilayer with varying distance using MACE-H and DeepH-E3. The onsite response error for the 2$\times$2$\times$1 shifted (\textbf{e}) and twisted (\textbf{f}) supercell counterparts. The onsite block components plotted in the figure are of different layers from the perturbed atoms.}
    \label{fig: locality}
\end{figure*}

So far, MACE-H has only been assessed for its performance on randomly held-out test set data of rattled and shifted monolayers and bilayers of two-dimensional materials and bulk gold. Following the approach of \citet{DeepH-E3}, we test the out-of-distribution performance of the model by performing predictions on unseen twisted bilayer configurations with varying twist angles. For the five bilayer datasets (C/graphene: noSOC, Bi$_2$Te$_3$: noSOC and SOC, Bi: soc, Bi$_2$Se$_3$) for which twisted configurations exist, MACE-H ($\nu=3$, $L_{\mathrm{max}}=5$) performs either as well or slightly worse than DeepH-E3 (Supplementary Fig. 6a). This effect is most significant for the Bi$_2$Te$_3$ dataset (with and without SOC), featuring more complex atomic environments. As a comparison, the MAE for SOC Bi$_2$Te$_3$ increases from 0.37/0.48 meV for the shifted bilayers to 1.25/0.70 meV for the twisted bilayers using MACE-H/DeepH-E3. For Bi$_2$Se$_3$, MACE-H and DeepH-E3 appear to perform similarly; however, the Bi$_2$Se$_3$ dataset is three times larger than the other datasets, indicating that larger dataset sizes are able to counteract this effect. For the twisted configurations, the many-body expansion seems to either not provide an advantage or reduce the accuracy of the model slightly, as shown in Supplementary Figs. 6a, 7, and 8. However, we note that even a matrix element MAE of 3~meV still yields band structures and DOS in the vicinity of the Fermi level that are effectively indistinguishable from the DFT reference. This effect, therefore, is acceptable in terms of model performance, but it is worth investigating further.

To investigate the difference between MACE-H and DeepH-E3 for the prediction of twisted bilayers, we analyse the models using the response to a slightly displaced atom in Bi\textsubscript{2}Te\textsubscript{3} bilayer (see Fig.~\ref{fig: locality}a) with the response defined as the prediction difference between the perturbed and the pristine conformations. Since the onsite matrix blocks are more straightforward to study the message passing of MACE-H with many-body expansion, we calculate the response error of the models for the onsite matrix blocks, while the response error is defined as the block-wise MAE of the difference between the model-predicted and DFT-predicted responses.
In Fig.~\ref{fig: locality}b, MACE-H outperforms DeepH-E3 for the shifted bilayers regardless of distance (the unperturbed configuration index is 72-0 in the dataset). The overall onsite response MAEs for the shifted Bi\textsubscript{2}Te\textsubscript{3} bilayer are $9.6\times10^{-2}$ and 0.13 meV for MACE-H and DeepH-E3, respectively. In contrast, MACE-H shows higher error compared to DeepH-E3 (Fig.~\ref{fig: locality}b), particularly for matrix blocks further away from the perturbed atom (beyond 5 \AA{}). The overall onsite response MAEs for the twisted Bi\textsubscript{2}Te\textsubscript{3} bilayer are 0.13 and $9.8\times10^{-2}$ meV for MACE-H and DeepH-E3, respectively. To exclude periodic image interaction, we constructed a $2\times2\times1$ supercell of the shifted and twisted Bi\textsubscript{2}Te\textsubscript{3} bilayer. For the supercells (shifted and twisted), the differences between MACE-H and DeepH-E3 are in the noise of the model as the interactions decay towards large distance from the perturbed atom (Fig.~\ref{fig: locality}e and f: $4.5\times10^{-2}$ versus $4.8\times10^{-2}$ meV for the shifted supercell and $4.14\times10^{-2}$ versus $4.47\times10^{-2}$ meV for the twisted supercell). Further analysis of the absolute response magnitude (Fig.~\ref{fig: locality}d) shows that the decreased response error of MACE-H for the twisted supercell beyond 10 \AA{} is a result of the MACE-H response attenuating faster with distance. 
The response magnitudes, defined as the mean absolute value of the matrix block, are $6.8\times10^{-3}$ and $1.5\times10^{-2}$ for the interlayer onsite blocks with distance beyond 10 \AA{} using MACE-H and DeepH-E3, respectively. In summary, MACE-H shows a tendency for locality, i.e., it emphasises messages at closer distance and provides an improved representation of the short-range environment around atoms and bonds. This comes at the cost of reducing sensitivity to subtle structural changes further away from the central atom or bond. In the case of the twisted structures in the original unit cells, this leads to slightly larger errors. For the supercells, this effect is not significant and increased locality may even be an advantage. The analysis for the second twisted bilayer configuration in the dataset and its $2\times2\times1$ supercell also showed the same tendency (Supplementary Figs. 9 and 10). Our analysis also indicates that the interlayer onsite blocks showed higher sensitivity to different models than the intralayer onsite blocks (Supplementary Note 5.2).

To understand how the many-body expansion makes MACE-H more local and increases its sensitivity to the environment at close vicinity to the central atom or bond, we refer back to the many-body expression (Eq.~\ref{eqn: mbe}). Using the iterative node-wise tensor product of irreps, the effect of edge components with larger contributions (which are usually edges with shorter distances) in the aggregation process will be magnified, while the importance of those with lower contributions will be reduced. Supplementary Fig. 6b
shows that by increasing the correlation order $\nu$ and maximum degree of hidden states $L_{\mathrm{max}}$, the MACE-H prediction accuracy is increased for the shifted bilayers and reduced for the twisted bilayers. On the contrary, lower correlation order ($\nu$=1) and degree of hidden states ($L_{\mathrm{max}}$=1) leads to reduced accuracy for the shifted bilayers and higher accuracy for the twisted bilayers. The difference is statistically significant, and the influence is even more salient using the maximum MAE among the subblock elements as the measure (Supplementary Fig. 7). Supplementary Fig. 8 reports model performance on shifted and twisted Bi$_2$Te$_3$ as a function of correlation order $\nu$ and maximum hidden state degree $L_{\mathrm{max}}$, which further supports this observation. Moreover, the locality of the model is partially related to the choice of radial basis function and envelope function (Supplementary Tables 6, 7, and 9).

\subsection*{Modified Shift-and-Scale Operation}\label{sc: shift and scale}

\begin{figure}
    \centering
    \includegraphics[width=3.3in,keepaspectratio]{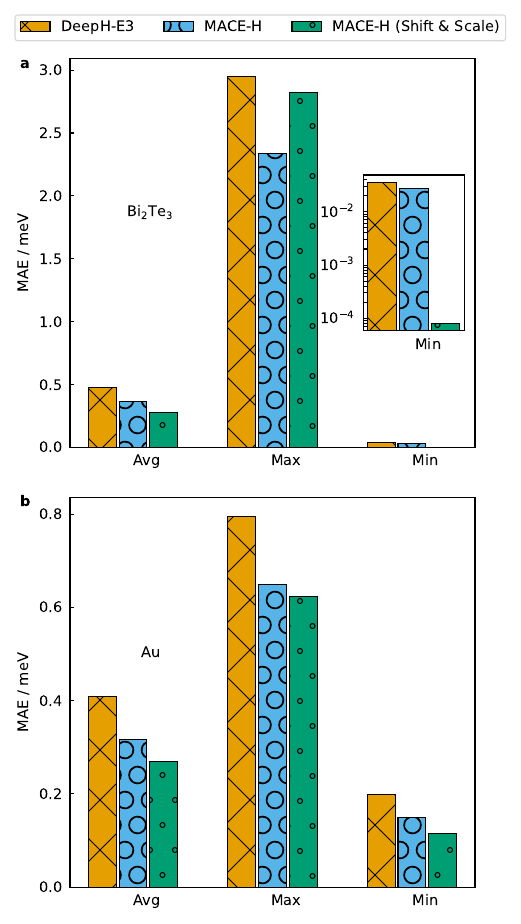}
    \caption{\textbf{Effect of Shift-and-Scale Operation on Output Features.} The average, maximum, and minimum MAE of matrix elements for MACE-H with/without scale-and-shift operation compared to DeepH-E3 for shifted Bi\textsubscript{2}Te\textsubscript{3} bilayer data (see \textbf{a}) and bulk Au (see \textbf{b}). The inset in panel \textbf{a} shows the magnified view of the minimum value of the error matrix. The statistical average, maximum, and minimum MAE values are obtained from the MAEs of the element pair-resolved orbital pairs.}
    \label{fig: shift and scale}
\end{figure}

In MLIPs, the graph neural network outputs are usually shifted and scaled according to the mean value and standard deviation of the species-aware energy labels in the dataset to fit the output energy. The shifting and scaling operation according to the distribution of the target labels will help the neural network to converge faster and achieve higher accuracy in the training process. In Hamiltonians, as values of matrix elements vary dramatically, the standard deviations within specific element-pair-resolved interaction blocks will also vary significantly. For instance, the standard deviation of different output irreps ranges from $5.42\times10^{-16}$ to 28.33 for the Bi$_2$Te$_3$ with SOC training data. The large magnitude difference between subblocks poses a challenge for conserving the norm of the model output while maintaining numerical stability. The direct implementation of shifting and scaling the model output incurred noticeable numerical instability, as shown in Supplementary Fig. 11. This is because irreps with larger standard deviations are prone to having larger mean square errors, as they are used in the loss function, and in the back-propagation process, the gradient will be further magnified by multiplying by the standard deviation, which tends to cause gradient explosion. Conversely, the irreps with lower standard deviations will experience vanishing gradients.

We rectify the gradient flow in MACE-H by removing scaling in the back propagation step and applying scaling only in the feed-forward process (see Supplementary Note 7.1). The general formula for the scaling and shifting operation is expressed as follows:
\begin{equation}
o_{ij,klm} = e_{ij,klm}^{(t)}\sigma_{Z_{ij}, kl} + \mu_{Z_{ij}, kl}
\end{equation}
where $\sigma_{Z_{ij}, kl}$ and $\mu_{Z_{ij}, kl}$ are the precomputed standard deviations and mean values of the corresponding target irreps. More specifically, the edgewise matrix blocks in the training data are first converted into the irreps of direct sum form with tensor product contraction, each irrep component within is normalized afterwards, and the standard deviation $\sigma_{Z_{ij}, kl}$ of each irrep component’s norm is calculated separately for each element-pair $Z_{ij}$. The mean values of the scalar components $\mu_{Z_{ij}, kl}$ are calculated for each element-pair $Z_{ij}$, while the mean values will be set to zero for non-scalar vectors. More detailed discussions can be found in Supplementary Note 7. 

The modified shift-and-scale operation in MACE-H improves the average and maximum test set MAE for both the shifted Bi\textsubscript{2}Te\textsubscript{3} bilayer with SOC and bulk Au datasets (Figure~\ref{fig: shift and scale} and Supplementary Fig.~12). In particular, the minimum MAE between the sub-block elements decreases by more than two orders of magnitude for the SOC Bi\textsubscript{2}Te\textsubscript{3} since values of coupling matrix elements between spin channels are significantly lower than values of spin-diagonal blocks. Above all, the resulting MAE for Bi\textsubscript{2}Te\textsubscript{3}
is 0.255 meV compared to 0.368 meV for MACE-H without the modified shift-and-scale implementation.

\subsection*{Label-Free Accuracy Estimation of the Predicted Matrices}

\begin{figure}
    \centering
    \includegraphics[width=3.3in,keepaspectratio]{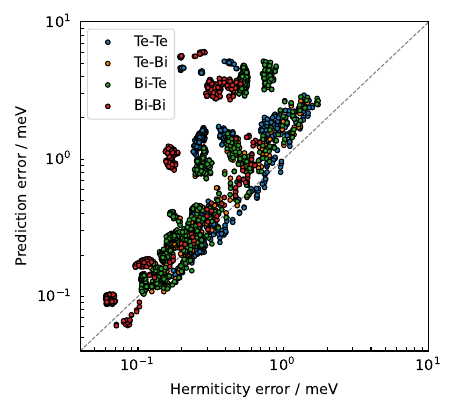}
    \caption{\textbf{Correlation of Predicted Matrix Block Accuracy with Hermiticity of the Predicted Matrix.} The correlation between the absolute value of matrix element prediction error and the hermiticity in the twisted Bi\textsubscript{2}Te\textsubscript{3} configuration resolved for different element-pairs.}
    \label{fig: uncertainty}
\end{figure}

The MACE-H model does not strictly impose hermiticity of the real-space KS Hamiltonian matrix, but it is rather imposed post-prediction through symmetrisation. The degree to which the model violates hermiticity can be evaluated by the self-adjointness of each atom pair:
\begin{equation}
    \mathrm{Error}_{\mathrm{herm}} = \left\| H_{ij,\bm{R}} - H_{ji,-\bm{R}}^{\dag} \right\|
\end{equation}
where $H_{ij, \bm{R}}$ is the real-space Hamiltonian block between $i$ and $j$ with lattice translation vector $\bm{R}$, and $H_{ji, -\bm{R}}^{\dag}$ is the conjugate transpose of the real-space Hamiltonian block. 

Fig. \ref{fig: uncertainty} compares the block-wise prediction error to the hermiticity error, which shows a positive correlation. Subblocks with high prediction error are also likely to violate hermiticity. We can consider the prediction of the two hermitian conjugate blocks of the matrix as an in-built ensemble of two, as the two subblocks will be exposed to different stochastically optimised parameter blocks in the model. As such, the degree by which hermiticity is violated could serve as a computationally efficient metric in active learning workflows to identify blocks with large errors without recourse to DFT ground truth labels. Furthermore, the comparison of the median numbers of prediction errors for matrix blocks with different quantiles of hermiticity error is given in Supplementary Fig.~13. We further compared prediction errors and hermiticity errors across different configurations using a series of models trained with varying training data sizes on our Au dataset (Supplementary Fig.~14). The positive correlation between prediction error and hermiticity error is consistently observed.

\section*{Discussion}

In this work, we propose the MACE-H model, which combines many-body message passing with an equivariant graph neural network for predicting the DFT Kohn-Sham Hamiltonian matrix and, subsequently, electronic properties such as the band structure and electronic DOS. This model is generally suitable for extended materials with diverse chemical composition, as is shown here for a range of two-dimensional and bulk materials. The model achieves ab-initio-level accuracy across a range of materials datasets. The use of many-body messages and a node degree expansion block improves the prediction accuracy compared to GNNs that employ strictly two-body messages, such as DeepH-E3. Model validation using eigenvalue error and electronic entropy error metrics also confirmed that the band structure derived from the predicted real-space KS Hamiltonian matrices is on par with DFT calculations. The use of many-body message passing leads to increased data efficiency of the models at no significant computational overhead for inference or training. We furthermore implement a modified shift-and-scale operation on individual output irreducible representations of the model that improves numerical stability and reduces noise in matrix subblocks with small values. Finally, the degree with which the model satisfies hermiticity can be used as an approximate gauge of the accuracy of different subblocks, as it is found to correlate with the prediction errors. 

Furthermore, we also observed that higher body-order message passing emphasizes local over distant messages, leading to a better representation of the local chemical environment around the central atom or bond where dominant Hamiltonian interactions occur. As a result, it is more data-efficient and more accurate. In turn, however, there is less sensitivity for configurational changes at larger distances, as seen for predictions on twisted bilayers of bismuth telluride. The behaviour observed for twisted bilayers, which show subtle configurational differences at large range, may be remedied in the future by partitioning the orbital interactions via the potential into short- and long-range components and capturing these with messages of different body order.

For the case of bulk gold predictions, we show that the MACE-H model is compatible with data generated by all-electron DFT packages such as FHI-aims. Firstly, KS DFT Hamiltonians can be pre-processed with a core projection to generate valence-only Hamiltonians that serve as training data. Secondly, accurate atomic orbital basis evaluations require a large cutoff radius that will lead to low sparsity in the Hamiltonian and matrix subblocks that span many orders of magnitude. Subblocks with very small values can lead to stability issues and background noise. To this end, sparsity can be enforced by truncation of matrix values below a threshold, and a modified shift-and-scale operation ensures smooth and stable training.

The model is already fully integrated into the DeepH data pipeline and is able to train directly on FHI-aims output data \citep{fhiaimsroadmap}. In the future, we plan to further improve this integration to harness the highly optimised parallel matrix algebra in FHI-aims and other codes for the diagonalisation of large KS Hamiltonian matrices, which forms the natural bottleneck in electronic structure predictions. The model can also be applied to the prediction of the density matrix in local orbital product basis representation, which offers the possibility to accelerate self-consistent-field convergence \citep{schnorb, zhangDM}. As such, MACE-H offers a promising route to reduce bottlenecks in electronic structure algorithmss by augmentation with data-driven models.  

Even when a deep integration with electronic structure software is achieved for Hamiltonian prediction message passing models, further implementation improvements will be required to improve the scalability and parallelisation as well as the memory utilisation of the model during training and prediction. This will unlock new electronic structure prediction capabilities for dynamics and high-throughput materials discovery as well as improve the transferability of materials surrogate models by retaining electronic structure information. Furthermore, developing rigorous, calibrated uncertainty quantification for equivariant Hamiltonian prediction will be necessary to implement active learning for improved data utilization and reduced computation overhead for data generation. Machine-learned Hamiltonian surrogates with high error and uncertainty control will fill a crucial gap in the computational materials prediction portfolio of the future by accelerating the prediction of transport properties, enabling embedded quantum dynamics and spectroscopy calculations.

\section*{Methods}

\subsection*{Model details}\label{sec:model_details}

In this work, Hamiltonians from OpenMX \citep{OpenMx1, OpenMx2} and FHI-aims \cite{blumInitioMolecularSimulations2009} electronic structure codes were used. They employ numeric atom-centered orbitals (NAOs) that have a well-defined cutoff radius beyond which the orbital is equal to zero. Therefore, the connectivity graph of an input structure during training in the model was defined according to the sparsity of the Hamiltonian (Supplementary Table 11). During inference, the sparsity was inferred from the overlap matrix which has the same sparsity as the Hamiltonian.

The connectivity graph defines for which atom pairs the Hamiltonian prediction is being performed, and also defines the atomic neighbourhoods $\mathcal{N}_{i}$ used in message passing (Eq.~\ref{eqn: Geo TP}).

The MACE-H model contains two radial embeddings (Fig.~\ref{fig: workflow}). In the node update, which consists of MACE and node degree expansion blocks, the radial distance $|\bm{r}_{ji}|$ between atoms $i$ and $j$ was embedded to a basis $e_{\mathrm{N}}(|\bm{r}_{ji}|)$. Bessel and Gaussian basis sets were used in this work (Supplementary Table 3). The other radial embedding $e_{\mathrm{E}}(|\bm{r}_{ji}|)$ was used in the edge update, where Gaussian basis were employed for all the datasets.

\subsection*{Training data}\label{sec:data_generation}

The Hamiltonian data for the two-dimensional materials are calculated using OpenMX \citep{OpenMx1, OpenMx2} and taken from Refs.~\citep{DeepH, DeepH-E3}. The monolayer graphene and MoS\textsubscript{2} configurations are generated from the ab initio molecular dynamics trajectory using the Vienna ab initio simulation package (VASP) \citep{VASP} with PBE functional \citep{PBE} and the projector-augmented wave (PAW) potential \citep{PAW1, PAW2}, containing 450 and 500 supercells, respectively. The shifted bilayer graphene, bismuthene, Bi\textsubscript{2}Se\textsubscript{3}, and Bi\textsubscript{2}Te\textsubscript{3} are constructed by shifting two relaxed neighboring van der Waals layers horizontally and imposing random perturbations below 0.1 {\AA} along each direction, containing 300, 576, 576, and 256 geometries, respectively. Notably, the Bi\textsubscript{2}Te\textsubscript{3} dataset contains two different versions, one with SOC and one without SOC. As a hold out assessment, twisted bilayers are constructed by twisting between different layers for graphene, bismuthene, Bi\textsubscript{2}Se\textsubscript{3}, and Bi\textsubscript{2}Te\textsubscript{3}, containing 9, 4, 2, and 2 geometries, respectively. More detailed information on the 2D material datasets is summarised in Supplementary Table 10 and can also be found in Ref. \citep{DeepH, DeepH-E3}. The overlap matrices used for the band structure calculation of Bi\textsubscript{2}Te\textsubscript{3} are calculated by OpenMX (version 3.9) with Bi8.0-s3p2d2 and Te7.0-s3p2d2 localized pseudo-atomic orbital (PAO) basis. The hyperparameter configuration files for the datasets can be found on Zenodo \citep{zenodo_dataset}.

For the locality analysis of Bi\textsubscript{2}Te\textsubscript{3}, configuration "72\_0" was used as a representative structure from the test set containing shifted structures and configurations "1\mbox{-}2" and "1\mbox{-}3" were taken to represent twisted bilayer structures. The Hamiltonians of unperturbed and perturbed unit cells and $2\times2\times1$ supercells of these configurations were calculated with OpenMX (version 3.9) using the same settings as described above. The perturbed atom is subjected to displacement by 0.1 \AA{} in three directions simultaneously. 

An unlabelled dataset of 1000 Au(FCC) $4\times4\times4$ supercell configurations was generated by performing Langevin molecular dynamics (MD) simulations with an Effective Medium Theory (EMT) potential provided in ASE 3.22.1 \cite{larsenAtomicSimulationEnvironment2017} at 1000~K for 100~ps and saving frames every 100~fs. From these configurations, the 200 most diverse datapoints were selected by Farthest Point Sampling (FPS) with configuration-averaged ACE descriptors \cite{qiPointNetDeepHierarchical2017,drautzAtomicClusterExpansion2019,dussonAtomicClusterExpansion2022}. Ten configurations were randomly selected for testing, and the remaining 190 configurations we randomly split into training and validation sets with a ratio of 4:1.

Hamiltonian and overlap matrices were obtained by performing DFT calculations with the FHI\mbox{-}aims electronic structure code (version 240926, commit df50e6022) \cite{blumInitioMolecularSimulations2009} using the PBE functional \cite{PBE} and the frozen-core approximation \cite{yuAccurateFrozenCore2021}. FHI\mbox{-}aims provides a range of numerical atomic basis sets and real-space integration grid settings for each chemical element. For more information, see \cite{blumInitioMolecularSimulations2009}. A customised ‘tight’ basis set was used for gold, which was modified by removing a subset of extra valence orbitals, resulting in a \textit{tier1-sdfgh} basis set with a 6 Å radial cutoff. Additionally, radial and angular densities of the real-space integration grid were increased beyond ‘tight’ settings. The calculations were performed with a $7\times7\times7$ k-point grid. Raw calculation data with further details on computational settings can be found on the NOMAD materials repository~\citep{nomad_dataset}.

The core states were projected from DFT data as outlined in Section \nameref{sec:core_projection}. Furthermore, all the Hamiltonian and overlap blocks with an edge distance above 10~Å were set to zero, even though the maximum edge distance in unprojected data was equal to 12~Å arising from the basis function radial cutoff of 6~Å in DFT calculations. We found that keeping the edges up to 12~Å significantly slowed down both training and inference, even though edges beyond 10~Å were found to only marginally affect Hamiltonian eigenvalues as discussed in the Supplementary Note 9. Furthermore, the performance of the model deteriorated for edges beyond 10~Å because such edges correspond to very small absolute Hamiltonian values, which pose a challenge for the current model architecture. We plan to modify the architecture of the model in the future to better tackle systems where truncation of long edges may not be justified.

Finally, the valence-only Au data was converted to a DeepH\mbox{-}compatible format. In this format, the matrices are stored in HDF5 files as key-value pairs containing interacting atoms and their corresponding matrix block, following OpenMX spherical harmonics convention. Additional files store other metadata of the system, such as angular momenta of the basis set and atom positions. The dataset, together with a more thorough explanation of the data format, can be found on Zenodo \cite{zenodo_dataset}.

\subsection*{Core-Orbital Projection}\label{sec:core_projection}

To speed up training and inference without sacrificing accuracy in the valence-energy region, the low-energy core states of gold were projected from the DFT data. This was achieved by first mapping the real-space Hamiltonian to reciprocal space on a dense k-point grid:
\begin{equation}\label{eq:fourier}
    \bm{H}_{\bm{k}} = \sum_{\bm{R}} \mathrm{e}^{-i\bm{k} \cdot \bm{R}}\, \bm{H}_{\bm{R}}\,,
\end{equation}
where $\bm{H}_{\bm{R}}$ is the real-space Hamiltonian between the orbitals in the central unit cell and the unit cell at $\bm{R}$. The reciprocal-space Hamiltonians were then partitioned into $\bm{H}_{\bm{k},\mathrm{cc}}$, $\bm{H}_{\bm{k},\mathrm{vv}}$, and $\bm{H}_{\bm{k},\mathrm{cv}}$, corresponding to core-core, valence-valence, and core-valence blocks, respectively. The same operation was done analogously using Eq.~\ref{eq:fourier} for the real-space overlap matrix to yield $\bm{S}_{\bm{k},\mathrm{cc}}$, $\bm{S}_{\bm{k},\mathrm{vv}}$, and $\bm{S}_{\bm{k},\mathrm{cv}}$. An equivariance-preserving linear basis transformation $\bar{\bm{B}}_{\bm{k}}$ was performed to minimise core-valence coupling \cite{laikovIntrinsicMinimalAtomic2011}:
\begin{align}\label{eq:laikov1}
    \bm{H}_{\bm{k}} = \begin{pmatrix}
        \bm{H}_{\bm{k},\mathrm{cc}}^{\vphantom{\dag}} & \bm{H}_{\bm{k},\mathrm{cv}}^{\vphantom{\dag}} \\
        \bm{H}_{\bm{k},\mathrm{cv}}^{\dag} & \bm{H}_{\bm{k},\mathrm{vv}}^{\vphantom{\dag}}
    \end{pmatrix}\quad \xrightarrow[
    \bar{\bm{H}}_{\bm{k}}\, =\, \bar{\bm{B}}_{\bm{k}}^{\dag}\, \bm{H}_{\bm{k}}^{\vphantom{\dag}}\, \bar{\bm{B}}_{\bm{k}}^{\vphantom{\dag}}
    ]{
    \bar{\bm{B}}\, =\, \left(\begin{smallmatrix}
        \bm{S}_{\bm{k},\mathrm{cc}}^{-\frac{1}{2}} & -\bm{S}_{\bm{k},\mathrm{cc}}^{-1\vphantom{\frac{1}{2}}} \bm{S}_{\bm{k},\mathrm{cv}}^{\vphantom{\frac{1}{2}}}\\
        \bm{0} & \bm{I}
    \end{smallmatrix}\right)
    }\quad
    \bar{\bm{H}}_{\bm{k}} \approx \begin{pmatrix}
        \bar{\bm{H}}_{\bm{k},\mathrm{cc}} & \bm{0} \\
        \bm{0} & \bar{\bm{H}}_{\bm{k},\mathrm{vv}}
    \end{pmatrix}\,,
\end{align}
where $\bm{I}$ is the identity matrix. The valence-only reciprocal-space Hamiltonians $\bar{\bm{H}}_{\bm{k},\mathrm{vv}}$ were mapped back to real space:
\begin{align}
    \bar{\bm{H}}_{\bm{R},\mathrm{vv}} = \frac{1}{N_{\bm{k}}} \sum_{\bm{k}} \mathrm{e}^{i\bm{k} \cdot \bm{R}}\, \bar{\bm{H}}_{\bm{k},\mathrm{vv}}\,.
\end{align}
For the gold dataset, it was chosen to only include 5d, 6s, and 6p orbitals in the valence partition, reducing the dimension of matrix blocks from $43 \times 43$ to $9 \times 9$. The resulting valence-only Hamiltonians $\bar{\bm{H}}_{\bm{R},\mathrm{vv}}$ were used as ground-truth labels for gold in this work.

The linear transformation $\bar{\bm{B}}_{\bm{k}}$ effectively changes the underlying basis set of the Hamiltonian. The transformed basis function in reciprocal space $\bar{\chi}_{\bm{k},(nlm)}(\bm{r} - \bm{t}_{i})$ with index $n$ and angular momentum and angular momentum projection quantum numbers $l$ and $m$, centered on atom $i$ (position $\bm{t}_{i}$), can be written as:
\begin{align}
    \bar{\chi}_{\bm{k},(nlm)}(\bm{r} - \bm{t}_{i}) = \sum_{jn'l'm'} \chi_{\bm{k},(n'l'm')}(\bm{r} - \bm{t}_{j})\, \bar{B}_{\bm{k},(jn'l'm'),(inlm)}\,,
\end{align}
where $\chi_{\bm{k},(n'l'm')}(\bm{r} - \bm{t}_{j})$ denotes an original basis function centered on atom $j$ (position $\bm{t}_{j}$) with an index $n'$ and angular momentum and angular momentum projection quantum numbers $l'$ and $m'$. More specifically, taking into account the explicit form of $\bar{\bm{B}}_{\bm{k}}$ in Eq.~\ref{eq:laikov1}, the transformed valence basis function can be written as:
\begin{align}\label{eq:transf_basisfunc_valence}
    \bar{\chi}_{\bm{k},(nlm)}^{(\mathrm{v})}(\bm{r} - \bm{t}_{i}) = \chi_{\bm{k},(nlm)}^{(\mathrm{v})}(\bm{r} - \bm{t}_{i}) - \sum_{\substack{jn'l'm'\\n'\in\, \mathcal{N}_{\mathrm{c}}}} \chi_{\bm{k},(n'l'm')}^{(\mathrm{c})}(\bm{r} - \bm{t}_{j}) \left[ \bm{S}_{\bm{k},\mathrm{cc}}^{-1} \bm{S}_{\bm{k},\mathrm{cv}}^{\vphantom{-1}} \right]_{(jn'l'm'),(inlm)}\,,
\end{align}
where $(\mathrm{v})$ and $(\mathrm{c})$ labels denote the valence and the core orbitals, respectively, and $\mathcal{N}_{\mathrm{c}}$ is a set of basis function indices that belong in the core orbital partition. This technically makes the Hamiltonian $\bar{\bm{H}}_{\bm{R},\mathrm{vv}}$ more difficult to learn because the valence basis functions themselves depend on the atomic environment due to the presence of the overlap matrix in Eq.~\ref{eq:transf_basisfunc_valence}. Nonetheless, the perturbation to the valence orbitals is minimal as long as the starting core-valence overlap is small, and it only depends on the local atomic environment.

Although not strictly required for training the model, the same projection procedure was performed for overlap matrices to obtain the single-particle energies by solving the generalised eigenvalue problem:
\begin{align}
    \bar{\bm{H}}_{\bm{k},\mathrm{vv}}\bar{\bm{C}}_{\bm{k},\mathrm{vv}} = \bar{\bm{S}}_{\bm{k},\mathrm{vv}}\bar{\bm{C}}_{\bm{k},\mathrm{vv}} \bm{\epsilon}_{\bm{k},\mathrm{vv}}\,,
\end{align}
where $\bar{\bm{H}}_{\bm{k},\mathrm{vv}}$ and $\bar{\bm{S}}_{\bm{k},\mathrm{vv}}$ correspond to valence-only reciprocal-space Hamiltonian and overlap matrices obtained by mapping $\bar{\bm{H}}_{\bm{R},\mathrm{vv}}$ and $\bar{\bm{S}}_{\bm{R},\mathrm{vv}}$ to reciprocal space (Eq.~\ref{eq:fourier}). $\bar{\bm{C}}_{\bm{k},\mathrm{vv}}$ is a matrix containing eigenvectors in each column and $\bm{\epsilon}_{\bm{k},\mathrm{vv}}$ corresponds to a matrix which contains the resulting valence-only eigenvalues on the diagonal.

Further details, such as the effect of core projection on valence eigenvalues and the projection convergence, can be found in Supplementary Note 9.

\subsection*{Accuracy Metrics}\label{sec:metrics}

In order to investigate model performance beyond Hamiltonian MAE, two additional physically inspired error metrics based on Hamiltonian eigenvalues were used.

To quantify the accuracy in the occupied eigenvalue region by ignoring less physically relevant high-energy states, a smeared eigenvalue error was defined as:
\begin{gather}
    \Delta{\epsilon}(T) = \frac{\sum_{\bm{k}i} | \epsilon_{\bm{k}i} - \tilde{\epsilon}_{\bm{k}i} |\, w_{\bm{k}} f(\epsilon_{\bm{k}i}, \mu, T)}{\sum_{\bm{k}i} w_{\bm{k}} f(\epsilon_{\bm{k}i}, \mu, T)}\,,
\end{gather}
where $w_{\bm{k}}$ is the weight of the k-point $\bm{k}$ on a dense k-point grid, and $\epsilon_{\bm{k}i}$ and $\tilde{\epsilon}_{\bm{k}i}$ denote DFT and predicted eigenvalues of the electronic band $i$, respectively. The denominator ensures the error is normalised with respect to the number of electrons in the unit cell. The chemical potential $\mu(T)$ required for the Fermi-Dirac distribution $f(E,\mu,T) = \left(1 + \exp{ 
\left[ \left(E-\mu \right)/\,k_{\mathrm{B}}T \right]}\right)^{-1}$ at a given temperature $T$ was obtained by enforcing charge neutrality via:
\begin{equation}
    N_{\mathrm{el}} = \int_{-\infty}^{+\infty} \mathrm{DOS}(E)\, f(E, \mu, T) \,\mathrm{d}E\,,
\end{equation}
where $N_{\mathrm{el}}$ is the number of electrons in the unit cell, $k_{\mathrm{B}}$ is the Boltzmann constant, and the same $\mathrm{DOS}(E) = \frac{1}{V}\sum_{\bm{k}i} \delta(E - \epsilon_{\bm{k}i})$ was used at any chosen temperature, with $V$ being the volume of the unit cell.

To quantify how well the models predict eigenvalues close to the Fermi level, it was chosen to compare DFT and predicted electronic entropy densities:
\begin{equation}
    s(T) = -k_{\mathrm{B}}\int_{-\infty}^{+\infty} \mathrm{DOS}(E)\, \left[f\ln{f} + (1 - f)\ln{(1 - f)}\right]\, \mathrm{d}E\,,
\end{equation}
where $f \equiv f(E, \mu, T)$ is the Fermi-Dirac distribution, and the $\mathrm{DOS}(E)$ is assumed to be temperature-independent as above. As before, chemical potentials for both DFT and predicted electronic entropies were estimated from charge neutrality. The electronic entropy error is reported as the absolute error between DFT and predicted values.

\section*{Data Availability}\label{sec: data availability}
The raw Au dataset is available in NOMAD (\url{https://doi.org/10.17172/NOMAD/2025.04.14-1}). The valence-only Au dataset, training configuration files to reproduce the models and the corresponding Python environment containers are available on Zenodo (\url{https://doi.org/10.5281/zenodo.15223696}). The 2D material datasets are available on Zenodo (\url{https://zenodo.org/records/7553640, https://doi.org/10.5281/zenodo.7553827, https://doi.org/10.5281/zenodo.7553843}) as released by ref. \citep{DeepH, DeepH-E3}.

\section*{Code Availability}
The MACE-H code in the current paper is available on GitHub: \url{github.com/maurergroup/MACE-H}.

\section*{Acknowledgements}
High-performance computing resources were provided via the Scientific Computing Research Technology Platform of the University of Warwick, the EPSRC-funded HPC Midlands+ computing centre for access to Sulis (EP/P020232/1), and the Northern Ireland High Performance Computing (NI-HPC) service for access to Kelvin2 (EP/T022175/1). The authors acknowledge funding through the UKRI Future Leaders Fellowship programme (MR/X023109/1), a UKRI frontier research grant (EP/X014088/1), and the EPSRC Centre for Doctoral Training in Modelling of Heterogeneous Systems (EP/S022848/1) at the University of Warwick.

\section*{Author Contributions}

J.R.K. and R.J.M. conceptualised and supervised the project. C.Q. developed the MACE-H model. V.V. generated data. V.V. and C.Q. prepared figures. All authors contributed to data analysis, interpretation, and manuscript writing.

\section*{Competing Interests}
Reinhard J. Maurer is an Associate Editor of npj Computational Materials. Reinhard J. Maurer was not involved in the journal’s review of, or decisions related to, this manuscript. The other authors do not have a competing interest.

\bibliography{references}

\end{document}


\maketitle

\beginsupplement

\clearpage
\makeatletter
\renewcommand{\numberline}[1]{#1\hspace{0.3em}}
\renewcommand{\numberline}[1]{Supplementary Note~#1\hspace{0.5em}}
\makeatother
\tableofcontents
\clearpage

\section{Hamiltonian Matrix MAE Analysis}
\begin{figure}[htbp!]
    \centering   \includegraphics[width=0.7\textwidth,height=0.7\textheight,keepaspectratio] {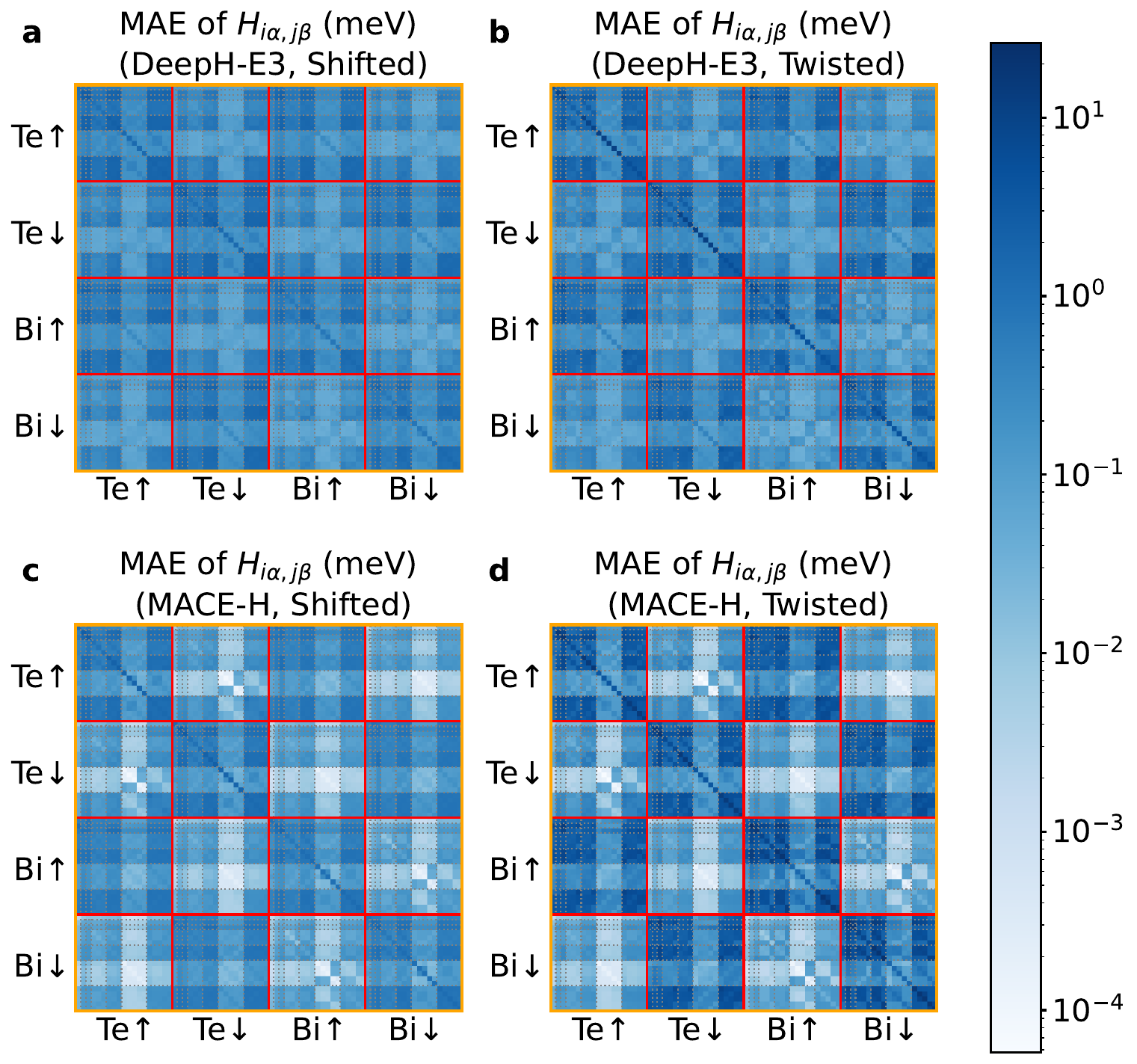}
    \caption{\textbf{Hamiltonian matrix element mean absolute error (MAE) comparison between MACE-H and DeepH-E3 for shifted and twisted Bi\textsubscript{2}Te\textsubscript{3}.} The MAE matrix of (\textbf{a}) shifted Bi\textsubscript{2}Te\textsubscript{3} bilayer predicted with DeepH-E3, (\textbf{b}) twisted Bi\textsubscript{2}Te\textsubscript{3} bilayer predicted with DeepH-E3, (\textbf{c}) shifted Bi\textsubscript{2}Te\textsubscript{3} bilayer predicted with MACE-H, (\textbf{d}) shifted Bi\textsubscript{2}Te\textsubscript{3} bilayer predicted with MACE-H. The MACE-H model in (\textbf{c}) and (\textbf{d}) includes shift-and-scale operations (see Supplementary Note S6).}
    \label{fig: MAE matrices}
\end{figure}

To compare the performance of MACE-H and DeepH-E3, we first trained a series of models on shifted Bi\textsubscript{2}Te\textsubscript{3} bilayers and then evaluated them using shifted test data and twisted bilayers. Supplementary Fig. \ref{fig: MAE matrices} shows the corresponding analytic MAE matrices. For performance on shifted bilayers, MACE-H outperformed DeepH-E3, with the average, maximum, and minimum of the error matrix being 0.28/0.48 meV, 2.82/2.95 meV and $7.98\times10^{-5}$/$3.59\times10^{-2}$ meV. In Supplementary Fig. \ref{fig: MAE matrices}a, the error matrix blocks (segmented by the red lines) between different spin channels are usually of similar magnitudes to the ones between the same spin channels for DeepH-E3. Supplementary Fig.~\ref{fig: MAE matrices}c shows that the MACE-H (employing shift and scale operation) manifested much lower errors between different spin channels. Moreover, the improvement of MACE-H is also manifested in the error matrix between the same spin channels. When applied to the twisted bilayers, the average, maximum, and minimum of the error matrices of MACE-H and DeepH-E3 are 1.31/0.70 meV, 26.11/11.11 meV and $5.78\times10^{-5}$/$2.59\times10^{-2}$ meV with noticeably larger errors of MACE-H, as shown in Supplementary Fig. \ref{fig: MAE matrices}b, d.

\section{Additional Gold Results}\label{sec:si_addgoldres}

\subsection{Correlation Order vs Number of Layers}\label{sec:si_addgoldres_corrvslayer}

In order to investigate the effect of increasing correlation order $\nu$ for the gold dataset, six different models were trained by varying the correlation order~$\nu$ and the number of message-passing layers~$T$ as shown in Supplementary Fig. \ref{fig:si_addgoldres_corrvslayer}. The model used in the main text corresponds to $\nu=2$ and $T=2$ unless stated otherwise.

\begin{figure}[htbp!]
    \centering
    \includegraphics[width=0.5\linewidth]{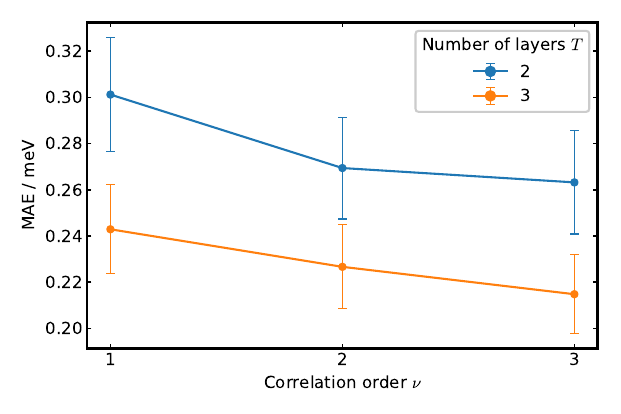}
    \caption{\textbf{Hamiltonian mean absolute error (MAE) for a range of models trained on the gold dataset with different correlation orders~$\nu$ and number of layers~$T$.} The error bars were obtained by computing the standard deviation~$\sigma_{\nu,T}$ of the MAE across configurations in the test set, and plotting $\pm \sigma_{\nu,T}$ intervals.}
    \label{fig:si_addgoldres_corrvslayer}
\end{figure}


\subsection{Learning Curves}\label{sec:si_addgoldres_lcurves}

To explore model convergence with respect to gold dataset size, MACE-H and DeepH-E3 models were trained with a range of training dataset sizes, and the MAE was computed for each model using the same test set as shown in Supplementary Fig. \ref{fig:si_addgoldres_lcurves}.

\begin{figure}[htbp!]
    \centering
    \includegraphics[width=0.5\linewidth]{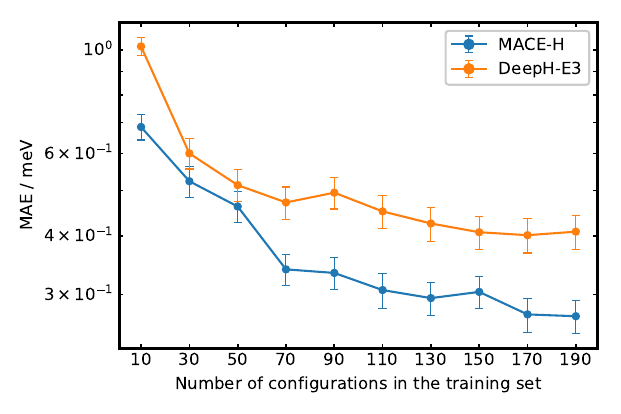}
    \caption{\textbf{Learning curves for MACE-H and DeepH-E3 for the gold dataset.} The reported error bars correspond to $\pm \sigma_{N_{\mathrm{train}}}$, where $\sigma_{N_{\mathrm{train}}}$ is the standard deviation of the MAE across configurations in the test set for each model trained on $N_{\mathrm{train}}$ configurations.}
    \label{fig:si_addgoldres_lcurves}
\end{figure}

\subsection{Time-to-Solution Comparison}\label{sec:si_gold_timetosolution}

In order to benefit from the computational speedup offered by the machine learning model, training data containing self-consistent Hamiltonians needs to be generated. For the machine learning model to be practical, the computational speedup offered by the model needs to outweigh the initial cost of generating the training data.

Bulk gold data consists of 200 self-consistent Hamiltonian matrices, each of which on average took 4 hours to calculate with 32 processes on an AMD EPYC 7742 CPU. This results in about 128 CPUhs per matrix, or 25600 CPUhs for the whole dataset.

On the other hand, test set prediction of 10 configurations with MACE-H took 10 seconds per configuration with 8 CPU threads on an AMD EPYC 7742 CPU. The prediction would be even faster on a GPU as shown for 2D systems in the main manuscript (Fig. 5). The computational time taken per configuration is equal to 0.02 CPUh, which is three orders of magnitude faster than DFT.

For larger configurations, we could expect even larger speedup offered by the machine learning model. This is because self-consistency cycles in DFT scale as $O(N_{\mathrm{at}}^{3})$, whereas MACE-H prediction scales linearly with the number of atoms. However, this speedup would diminish if the predicted Hamiltonian was diagonalised using a dense eigensolver with an $O(N_{\mathrm{at}}^{3})$ scaling.

\section{Benchmarking for molecular datasets} \label{benchmarking: molecule}

Apart from the model assessment for the 2D material and bulk gold datasets, we also evaluated our MACE-H performance on the MD17 molecular datasets, including water, ethanol, malondialdehyde, and uracil. To make a fair comparison with previous models, we adopted the same training, validation, and test data split. While $MSE$ loss is adopted for the 2D materials and bulk gold dataset, we employ the $loss=MAE+MSE$ for molecules, as also used by QHNet. The results on test sets using the MAE metric are shown in Supplementary Table \ref{table: molecule_datasets}. We can tell that the irreps-based equivariant GNNs (PhiSNet, QHNet, SPHNet, and our MACE-H) achieve higher accuracy than the covariant GNN (SchNorb) and the local-coordinate-based GNN (DeepH). Compared with these previous models, which employ 5 layers, our model requires only 3 convolution layers to achieve similar performance. The overall performance of these models is not only dependent on the model framework itself, but also sensitive to the choice of hyperparameters like learning rate or number of training steps. For example, in the existing literature, two different groups of results can be found for PhiSNet and QHNet (as in Supplementary Table \ref{table: molecule_datasets}), one using 200,000 steps for affordable training and the other adopting up to 1,000,000 steps for higher accuracy but with a much slower training process. Our MACE-H model strikes a good balance between the optimizing steps required and accuracy. The corresponding hyperparameters for the MACE-H model training of MD17 datasets (using the shift-and-scale operation) are in Supplementary Table \ref{table: hyperparameters_molecules} .

\begin{table}[htbp!]
\centering
\caption{MAE (in meV) of matrix elements on MD17 in comparison with other models. For the PhiSNet and QHNet, 5 convolutional layers are used, while MACE-H only uses 3 layers. Due to the high training overhead of TPs, previous literature reports two groups of results for PhiNet and QHNet, one using 200,000 steps for affordable training (the value before the slash symbol) and the other adopting up to 1,000,000 steps for higher accuracy but with a much slower training process (the value after the slash symbol).}
\small
\begin{tabular}{c c c c c c c} 
\toprule
Dataset & SchNorb & DeepH & PhiSNet & QHNet & SPHNet & MACE-H (ours) \\
\midrule
Water & 4.501 & 1.048 & 0.426/0.479 & 0.282/0.294 & 0.631 & 0.512 \\
Ethanol & 5.099 & 0.601 & 0.331/0.547 & 0.348/0.569 & 0.572 & 0.412 \\
Malondialdehyde & 5.200 & 0.547 & 0.335/0.580 & 0.326/0.562 & 0.586 & 0.349 \\
Uracil & 6.199 & 0.470 & 0.292/0.507 & 0.271/0.547 & 0.526 & 0.403 \\
\bottomrule
\end{tabular}
\label{table: molecule_datasets}
\end{table}

\begin{table}[htbp!]
\centering
\caption{Hyperparameters of the MACE-H models for the MD17 datasets. The $e_{\mathrm{N}}$, $l_r$, $N_B$, $N_{\mathrm{layers}}$, $\nu$, $L_{\mathrm{MACE}}$, $L_{\mathrm{node}}$, $L_{\mathrm{hidden}}$, and $L_{\mathrm{edge}}$ stand for the node-update radial basis (B for Bessel and G for Gaussian), learning rate, batch size, number of layers, correlation order, the largest azimuthal number of the intermediate MACE block irreps, the intermediate node-wise block irreps, the many-body expansion hidden states irreps, and the intermediate edge-wise block irreps.}
\resizebox{\textwidth}{!}{
\begin{tabular}{c c c c c c c c c c c} 
\toprule
Dataset & Train/val/test data split & $e_{\mathrm{N}}$ & $l_r$ & $N_B$ & epochs & $N_{\mathrm{layers}}$ & $\nu$ & $L_{\mathrm{MACE}}$ & $L_{\mathrm{hidden}}$ & $L_{\mathrm{edge}}$ \\
\midrule
Water & 500/500/3900 & G & 0.01 & 10 & 2000 & 3 & 3 & 4 & 4 & 4 \\
Ethanol & 25000/500/4500 & G & 0.01 & 64 & 1200 & 3 & 3 & 4 & 4 & 4 \\
Malondialdehyde & 25000/500/1478 & G & 0.01 & 64 & 1000 & 3 & 3 & 4 & 4 & 4 \\
Uracil & 25000/500/4500 & G & 0.01 & 64 & 1000 & 3 & 3 & 4 & 4 &  4 \\
\bottomrule
\end{tabular}
}
\label{table: hyperparameters_molecules}
\end{table}

\section{Data Efficiency of the Model} \label{sec: data-efficiency}

To assess the effect of different hyperparameter settings on MACE-H data efficiency, we adopted different settings. In Supplementary  Fig.~\ref{fig: data efficiency},  Deep-H uses spherical harmonics with $l$ up to 4, MACE-H uses the same settings in the edge-update block and specifies the node-wise MPNN block with spherical harmonics with $l$ up to 4, and hidden states with azimuthal number up to 2. The higher setting of MACE-H uses spherical harmonics $l$ up to 5 and hidden state azimuthal numbers up to 5. Models with two ($T=2$) and three layers are compared. The correlation order $\nu=3$ for all the MACE models. Benefiting from the higher expressive power of many-body expansion, MACE-H shows higher accuracy compared to DeepH-E3. While increasing the model depth helps to increase the data efficiency, the tighter setting of the many-body-expansion with larger correlation order $\nu$ and minimum azimuthal number $L_{\mathrm{max}}$ of hidden states seems to be more effective for higher accuracy across different training data sizes. The overall Bi\textsubscript{2}Te\textsubscript{3} dataset contains 256 configurations.

\begin{figure}[htbp!]
    \centering
    \includegraphics[width=0.4\textwidth,height=0.4\textheight,keepaspectratio]{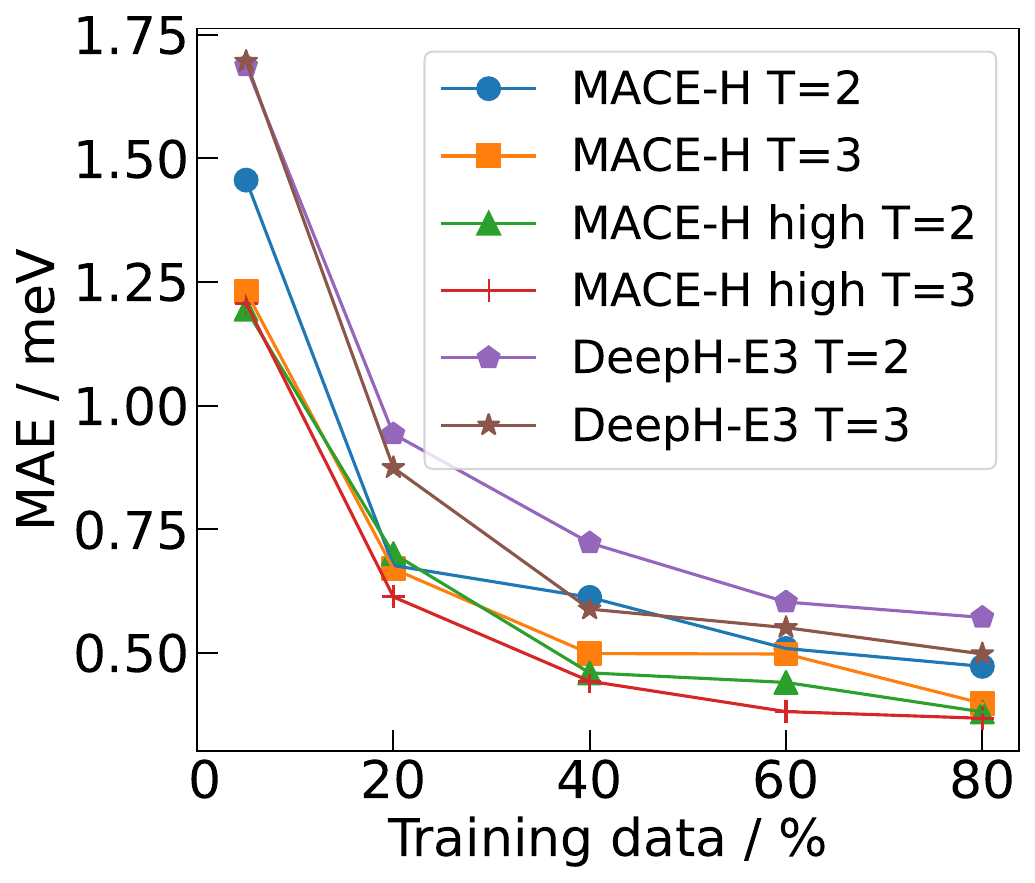}
    \caption{\textbf{Test set MAE for DeepH-E3 and MACE-H with different settings for shifted SOC Bi\textsubscript{2}Te\textsubscript{3} bilayers.} The label "high" means higher settings of spherical harmonics and the hidden states of many-body expansion, $T$, stand for model depth. The correlation order is $\nu=3$ for all the MACE-H models. The overall shifted SOC Bi\textsubscript{2}Te\textsubscript{3} bilayers dataset contains 256 configurations.}
    \label{fig: data efficiency}
\end{figure}

\section{Time-to-Solution Comparison: DeepH-E3 and MACE-H}

To compare the inference time between DeepH-E3 and MACE-H, we perform tests using systems of various sizes in Supplementary Fig.~\ref{fig: time-to-solution}. The inference time is obtained as time-to-solution using a single sample batch. Since the most computationally costly operation is the edge-wise update, the overall inference time of MACE-H is comparable to DeepH-E3 as they share the same edge-update block. No significant overhead due to the many-body expansion and the node degree expansion block can be identified.

\begin{figure}[htbp!]
    \centering    \includegraphics[width=0.7\textwidth,height=0.7\textheight,keepaspectratio]{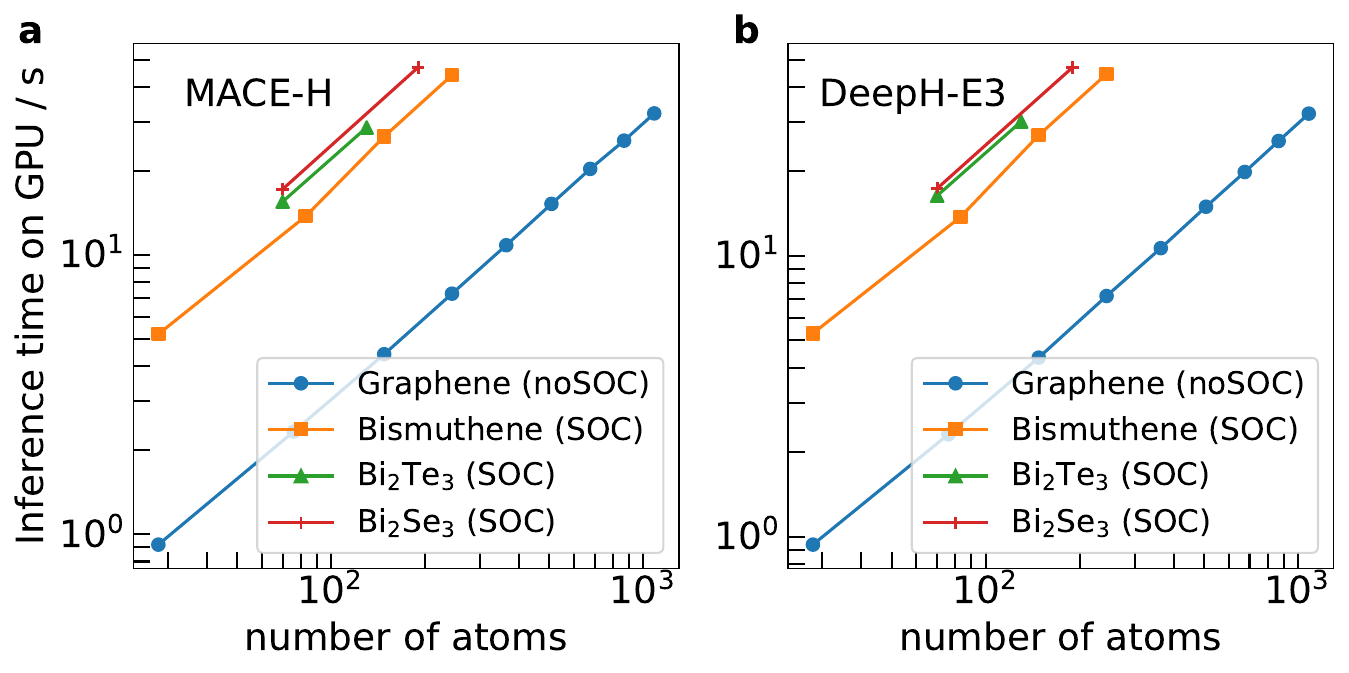}
    \caption{\textbf{Inference time comparison between (\textbf{a}) MACE-H and (\textbf{b}) DeepH-E3  on a single A100 GPU.}}
    \label{fig: time-to-solution}
\end{figure}

\section{Many-Body-Expansion for Different Interaction Range}

\subsection{The Influence of the Many-Body Expansion for Bilayer Predictions}

\begin{figure}[htbp!]
    \centering
    \includegraphics[width=0.85\textwidth,keepaspectratio]{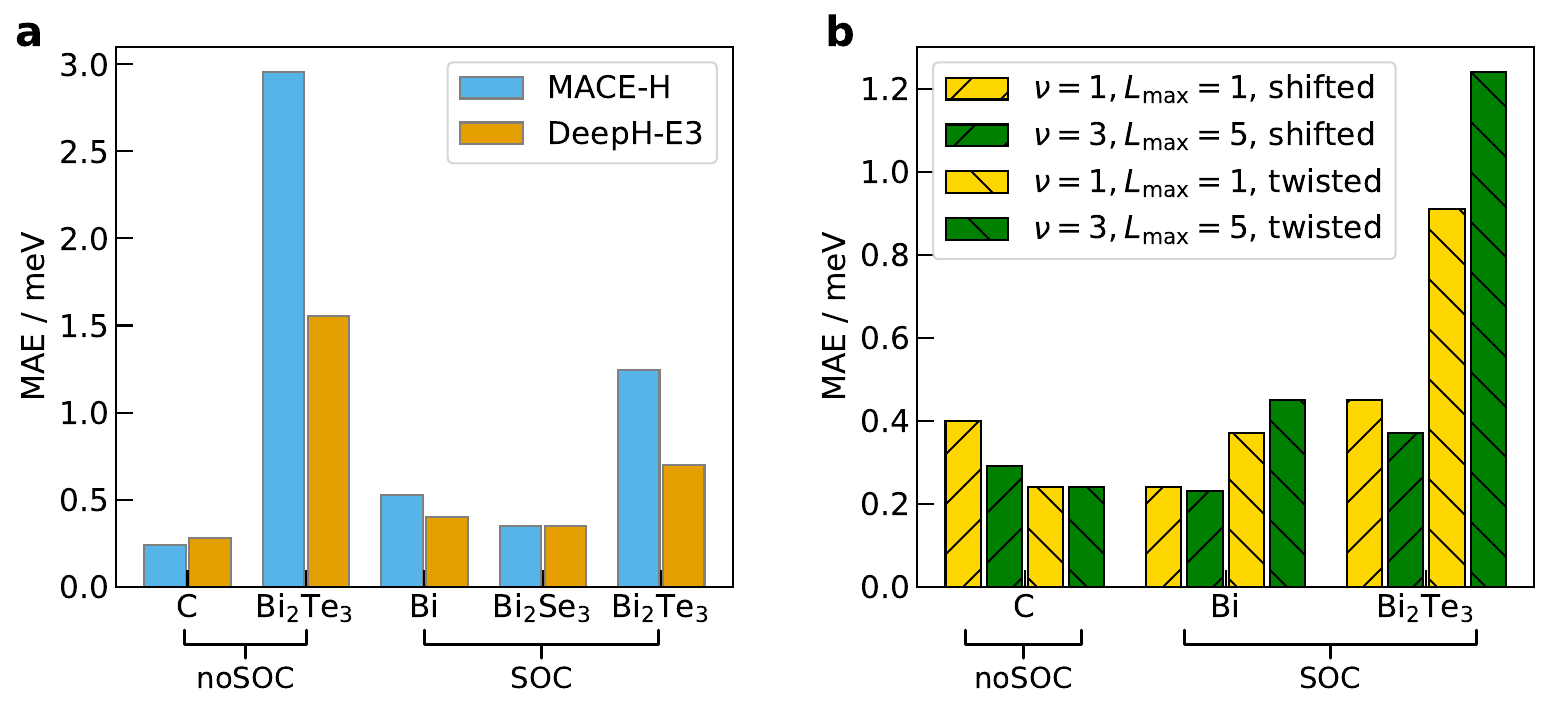}
    \caption{\textbf{Effect of correlation order for MACE-H predictions.} (\textbf{a}) Out-of-distribution MAE of matrix elements when applying the model trained on shifted bilayers to twisted bilayers for MACE-H and DeepH-E3. The correlation order $\nu=3$ is used for MACE-H. (\textbf{b}) Test set (shifted systems) and out-of-distribution (twisted systems) MAE of matrix elements using different settings of correlation order $\nu$ and $L_{\mathrm{max}}$ in the many-body expansion.}
    \label{fig: maceh_twsited}
\end{figure}

Supplementary Fig.~\ref{fig: maceh_twsited}a shows the performance of the model that was trained on shifted bilayers applied to twisted bilayer structures. While we found MACE-H predictions to be more accurate than DeepH-E3 for shifted bilayers, MACE-H either provides similarly accurate or slightly less accurate predictions for the twisted bilayers. We note that the Bi\textsubscript{2}Se\textsubscript{3} dataset is three times larger than Bi\textsubscript{2}Te\textsubscript{3}. Since shifted and twisted bilayers have similar atomic environments, the difference may arise from differences in the range of interactions. To study the influence of the many-body expansion on the shifted and twisted bilayers, we evaluate models with different settings of the many-body expansion block in Supplementary Fig.~\ref{fig: maceh_twsited}b. It should be noted that to make a fair comparison with DeepH-E3, the MACE-H model used here does not incorporate the shift-and-scale operation. It shows that a tighter setting (using larger correlation order $\nu$ and the maximal azimuthal number of hidden states $L_{\mathrm{max}}$) of the many-body expansion will favour the performance on shifted bilayers in the test set but will bring further error for twisted bilayers, indicating the locality preference. This effect is even more distinguishable for the maximum values of the Hamiltonian MAE matrices in Supplementary Fig.~\ref{fig: MBE range}, with the confidence interval showing the statistical significance. For predicting twisted bilayers with shifted bilayers as training data, it is recommended to use lower settings of $\nu$ and $L_{\mathrm{max}}$, and a larger dataset.

\begin{figure}[htbp!]
    \centering    \includegraphics[width=0.85\textwidth,height=0.85\textheight,keepaspectratio] {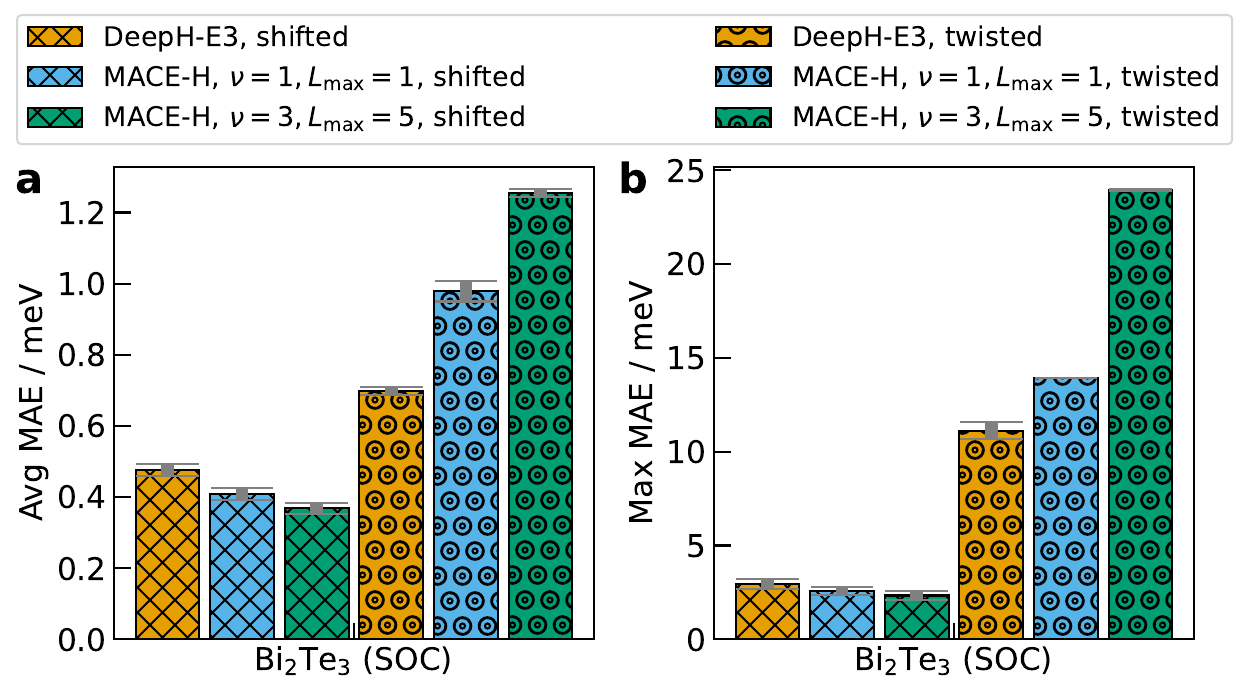}
    \caption{\textbf{(\textbf{a}) Average and (\textbf{b}) maximum values of the Hamiltonian error matrices.} Data is shown for shifted and twisted Bi\textsubscript{2}Te\textsubscript{3} bilayer using DeepH-E3 and MACE-H with different settings of correlation order $\nu$ and maximum azimuthal number of hidden state $L_{\mathrm{max}}$. The confidence interval of a standard deviation $\sigma_{\nu, L_{\mathrm{max}}}$ is taken across the error matrix metric of individual samples.}
    \label{fig: MBE range}
\end{figure}

\begin{figure}[htbp!]
    \centering
    \includegraphics[width=0.8\textwidth,height=0.8\textheight,keepaspectratio]{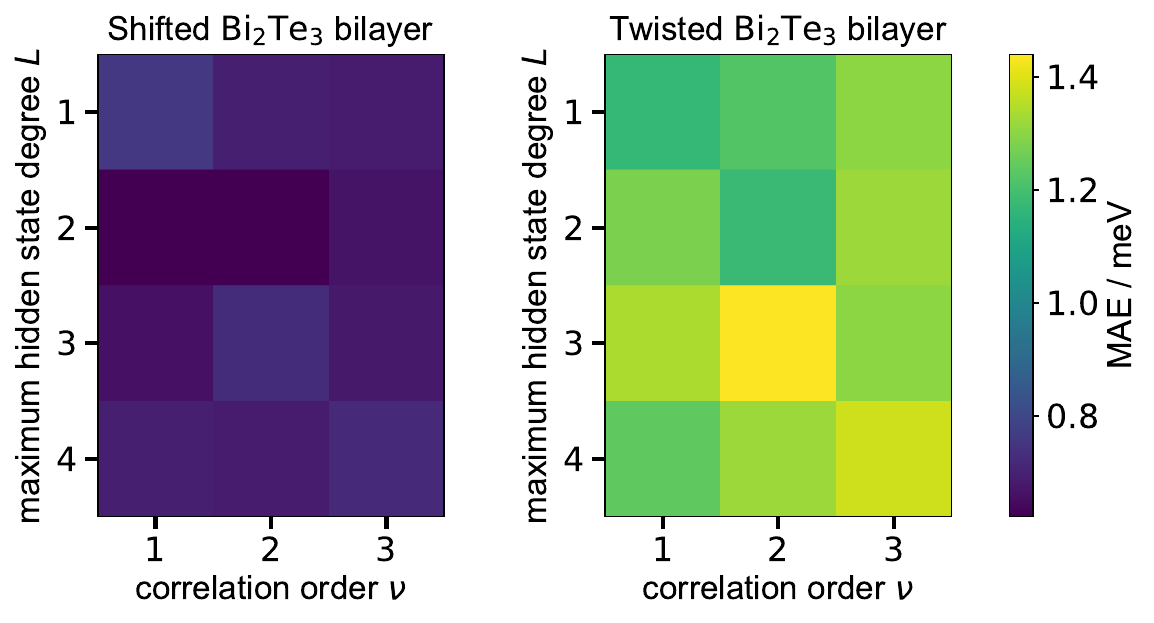}
    \caption{\textbf{The MAE for different settings of many-body expansion for shifted and twisted Bi\textsubscript{2}Te\textsubscript{3} bilayers.}}
    \label{fig: mbe}
\end{figure}

To study the effect of correlation order, $\nu$, and maximal azimuthal number of many-body expansion hidden states, $L_{\mathrm{max}}$, Fig. \ref{fig: mbe} shows the MAE using different hyperparameter settings. The other settings here are the same as those used in Supplementary Fig.~\ref{sec: data-efficiency}, i.e., we used spherical
harmonics with $l$ up to 4, hidden states with azimuthal number ($L_{\mathrm{max}}$) up to 2, and correlation order $\nu=3$. To avoid excessive computational overhead, we only used 20\% of the dataset as training samples, and no hyperparameter tuning was used with only fixed hyperparameters. The right panel shows the tendency for preferred locality with larger $\nu$ and $L_{\mathrm{max}}$.

\subsection{Atom Perturbation Analysis of Hamiltonian Model Response}

\begin{figure}[htbp!]
    \centering
    \includegraphics[width=\textwidth,keepaspectratio]{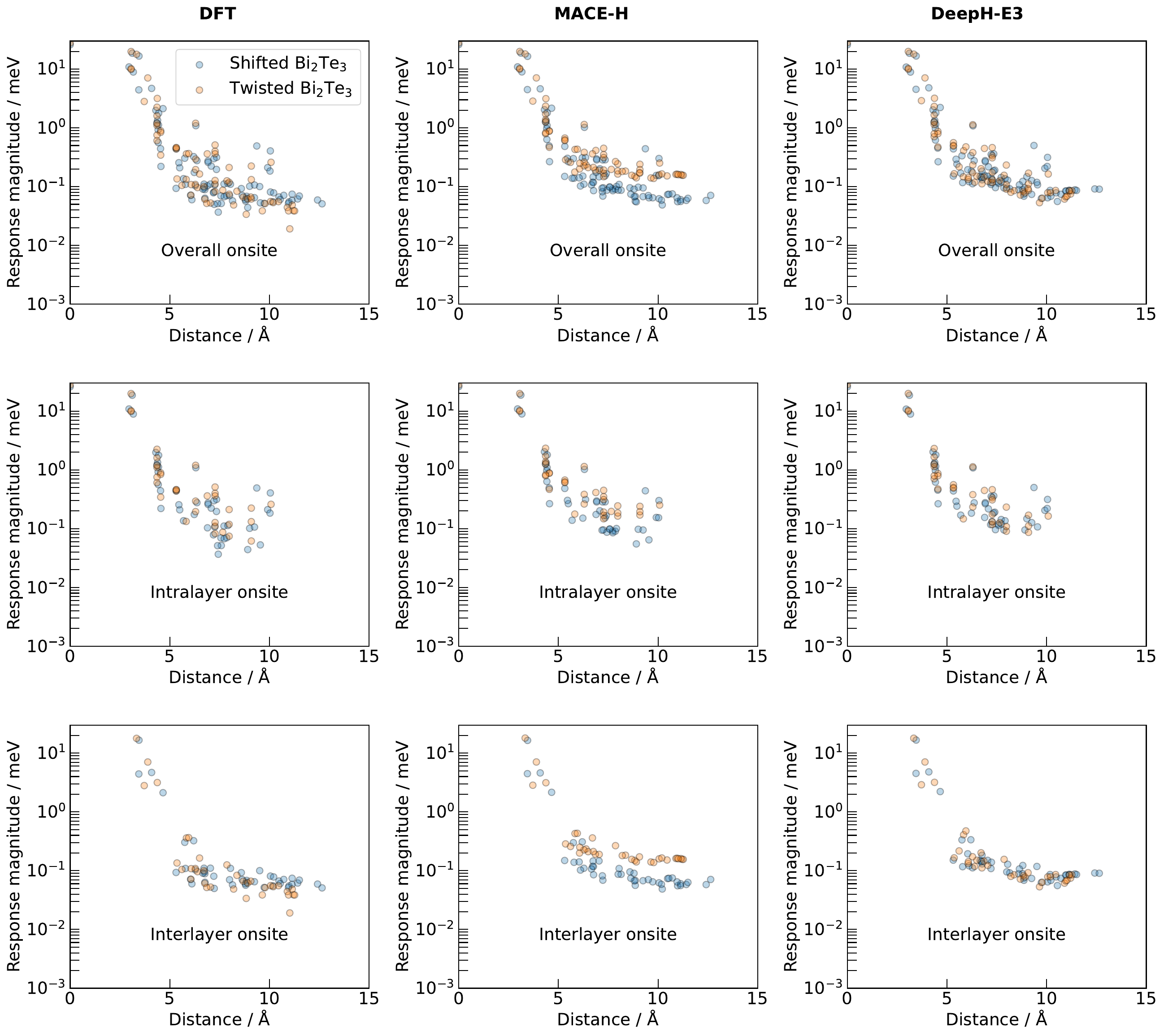}
    \caption{\textbf{Comparison of the predicted perturbation response magnitude between shifted (indexed by 72-0 in the dataset) and twisted (indexed by 1-2 in the dataset) unit cell bilayers using DFT, MACE-H, and DeepH-E3.}}
    \label{fig: response_magnitude_locality}
\end{figure}

To further study the interaction range captured by MACE-H, we analyse how MACE-H predicted on-site matrix blocks at varying distances from a single atom change upon the displacement of this atom. We studied the response magnitude by the perturbation for the shifted (indexed by 72-0 in the dataset) and twisted (indexed 1-2 and 1-3) Bi\textsubscript{2}Te\textsubscript{3} bilayer using DFT, MACE-H, and DeepH-E3. Supplementary Fig.~\ref{fig: response_magnitude_locality} shows the comparison between the shifted 72-0 and the twisted 1-2 unit cells using different methods. To further exclude the image interactions, we also construct the $2\times2\times1$ supercell counterparts of the unit cells, and the result also shows a similar observation. The data for the second twisted geometry shown in Supplementary Fig.~\ref{fig: response_error_locality} shows slightly different interaction range dependence of the models than the geometry shown in Figure 6 of the main manuscript, although the overall trends are the same.

For the shifted 71-0 bilayer unit cell, the overall onsite response error using MACE-H and DeepH-E3 are $9.6\times10^{-2}$ and $13.3\times10^{-2}$ meV, respectively. For the shifted 1-2 unit cell, the overall onsite response error using MACE-H and DeepH-E3 are $20.2\times10^{-2}$ and $13.8\times10^{-2}$ meV, respectively. However, for the supercell counterparts, the difference between MACE-H and DeepH-E3 regarding response error becomes negligible, with MACE-H showing a slightly lower error. We find that the change for supercells is mostly because MACE-H manifests lower response errors compared to DeepH-E3 for distances beyond 15 \AA{} in the twisted supercells, which is a result of faster attenuation of the predicted response magnitude for longer distance regions, also showing the locality. Nevertheless, the larger matrix MAEs of twisted bilayers for MACE-H also exist in the supercells.   

Here, the response to the perturbation $\Delta H_{ij}$ is defined as:
\begin{equation}
    \Delta H_{ij} =  H_{ij}^{\mathrm{perturbed}} - H_{ij}^{\mathrm{pristine}}
\end{equation}
where $H_{ij}^{\mathrm{perturbed}}$ and $H_{ij}^{\mathrm{pristine}}$ are the matrix blocks between atom $i$ and $j$ of the perturbed and pristine conformation. The response magnitude is defined as the mean absolute value of the response matrix elements:
\begin{equation}
    \left\|\Delta H_{ij}\right\| = \frac{1}{N_{orb}^{2}} \sum_{k_1=1}^{N_{\mathrm{orb}}} \sum_{k_2=1}^{N_{\mathrm{orb}}} \left| \Delta H_{ij} \right|_{k_1k_2}
\end{equation}
where $k_1$, $k_2$ indicate the indices of the matrix element, $N_{\mathrm{orb}}$ is the number of orbitals in a Hamiltonian matrix block. The response error is defined as:
\begin{equation}
    Error_{\Delta H_{ij}} = \frac{1}{N_{orb}^{2}} \sum_{k_1=1}^{N_{\mathrm{orb}}} \sum_{k_2=1}^{N_{\mathrm{orb}}} \left| \Delta H_{ij}^{\mathrm{pred}} - \Delta H_{ij}^{DFT} \right|_{k_1k_2}
\end{equation}
where $\Delta H_{ij}^{\mathrm{pred}}$ and $\Delta H_{ij}^{DFT}$ are the response matrices using machine learning model prediction and DFT calculation, respectively.

\begin{figure}[htbp!]
    \centering
    \includegraphics[width=0.9\textwidth,keepaspectratio]{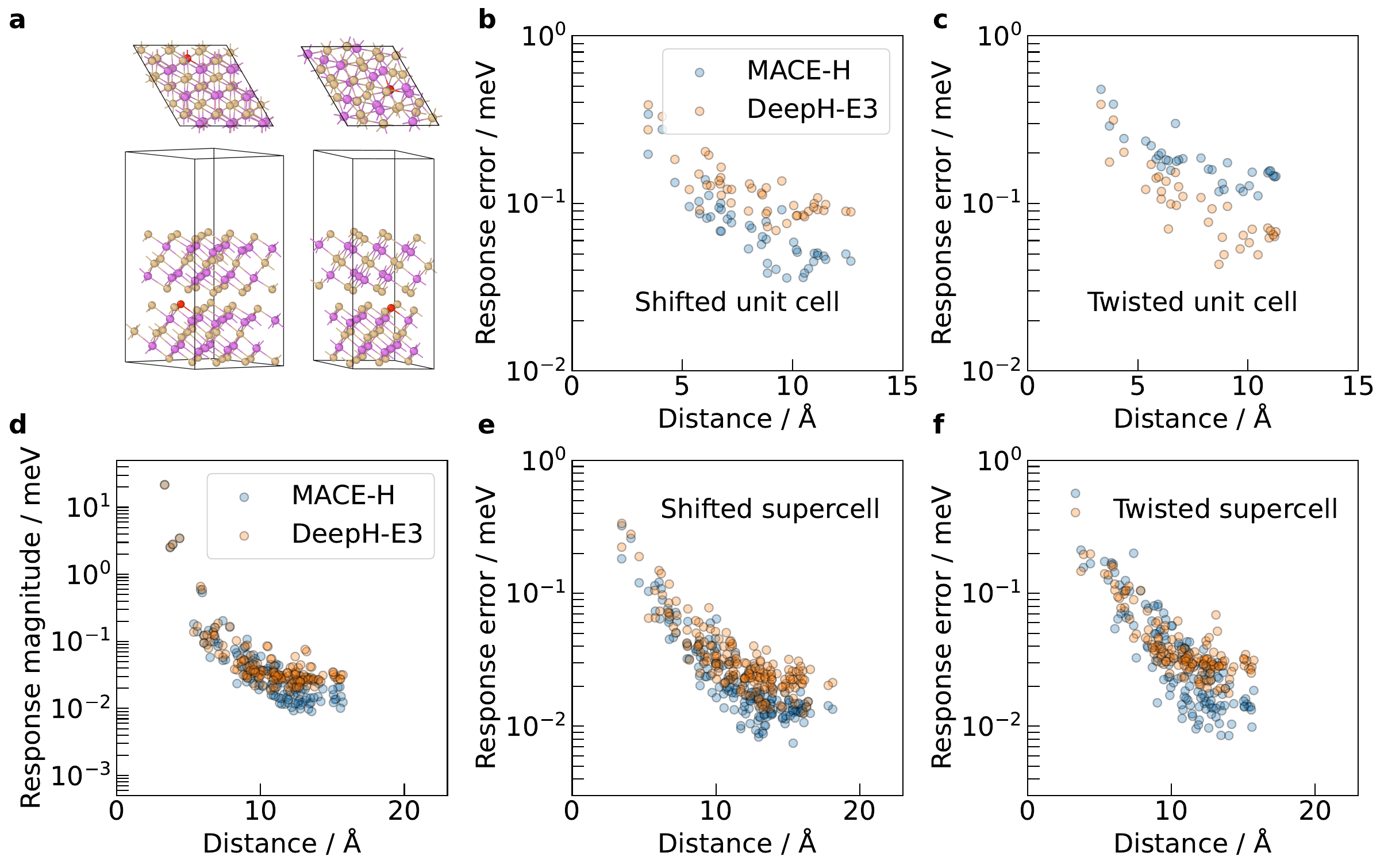}
    \caption{\textbf{Locality analysis of MACE-H compared to DeepH-E3 for the twisted bilayer based on single-atom displacement perturbation.} (\textbf{a}) The configuration of shifted (indexed by 72-0 in the dataset) and twisted (indexed by 1-2 in the dataset) Bi\textsubscript{2}Te\textsubscript{3} bilayer unit cells with the red atom being the perturbed Te atom. (\textbf{b}) The response error of onsite matrix blocks w.r.t. the distance from the perturbed atom for a shifted bilayer unit cell. (\textbf{c}) The onsite response error for a twisted bilayer unit cell. (\textbf{d}) The decay rate comparison for the 2$\times$2$\times$1 twisted bilayer with varying distance using MACE-H and DeepH-E3. The onsite response error for the 2$\times$2$\times$1 shifted (\textbf{e}) and twisted (\textbf{f}) supercell counterparts. The onsite block components plotted in the figure are of different layers from the perturbed atoms.}
    \label{fig: response_error_locality}
\end{figure}

\section{Shift-and-Scale Operation on the Last-Layer Irreps}

\subsection{Numerical Instability for Large Matrix Value Ranges}
\begin{figure}[htbp!]
    \centering\includegraphics[width=0.7\textwidth,height=0.7\textheight,keepaspectratio] {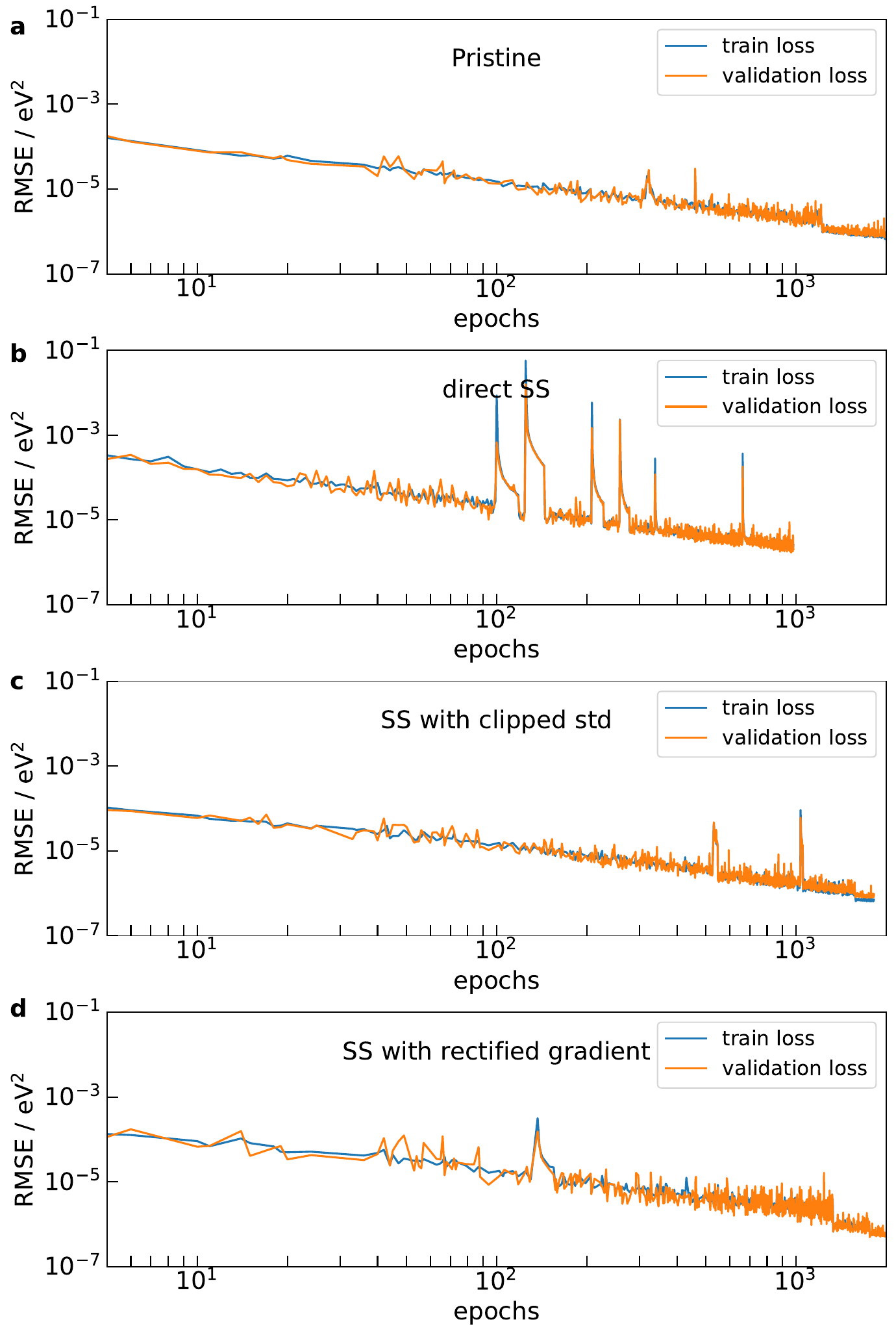}
    \caption{\textbf{The learning curve convergence of the models for shifted Bi\textsubscript{2}Te\textsubscript{3}.} (\textbf{a}) The original model training without shift-and-scale operation. (\textbf{b}) Direct application of the shift-and-scale operation to the last layer irreps. (\textbf{c}) Application of the shift-and-scale to the irreps with standard deviations of the norms lower than 1 manually set to 1. (\textbf{d}) Application of the shift-and-scale operation with a gradient bypass of the standard deviation in the backpropagation process.}
    \label{fig:SS_learningcurve}
\end{figure}

Due to the large norm difference between the irreps of different element-orbital-pair resolved sub-blocks, adjusting the output accordingly may have the potential to accelerate the model convergence and increase the accuracy. However, irreps for different sub-blocks exhibit enormous differences over many orders of magnitude, which is a much more challenging scenario than energy predictions in machine learning interatomic potentials. For example, Supplementary Fig. \ref{fig:SS_learningcurve}b shows that when we directly apply the shift-and-scale operation, the learning curve oscillates, the model fails to converge well, and the overall accuracy deteriorates. This is because as we scale the norms of the last layer irreps in the feed-forward process, it not only pushes the result closer to the magnitude of the target but also will scale the according gradients again in the backpropagation process, which will induce gradient explosion (ascribed to irreps with larger standard deviation) and disappearance (ascribed to the components with smaller standard deviation). Furthermore, the components with larger standard deviations will also have their errors magnified at early epochs after the scaling operation, which further contributes to the numerical instability. To tackle the issue, we first attempted to set the standard deviation components with magnitude lower than 1 to 1 (Supplementary Fig. \ref{fig:SS_learningcurve}c). This method alleviates the instability in the earlier stage, but the large oscillations persist in the later stage. Moreover, manually trimming the standard deviation with a threshold will not be suitable for a streamlined workflow since it is inconvenient to determine the optimal threshold. Thus, we adopted the method to allow the scaling operation in the feed-forward process but to sidestep the scaling factors in the gradient backpropagation process by redirecting the gradient flow:

\begin{equation}
o_{ij,klm} = o_{ij,klm}^{(t)} + (o_{ij,klm}^{(t)} (\sigma_{z_{i, j}, k} - 1)).detach() + \mu_{Z_{i,j}, k}
\end{equation}

where $\mu_{Z_{i,j}, k}$ and $\sigma_{z_{i, j}, k}$ stand for the precalculated standard deviations and mean values of the corresponding target irreps, and the mean values will be set to zero for non-scalar vectors, the detach means that the gradient will be truncated for that term. Although the learning curve still has one observable oscillation at earlier epochs, the later stage converges faster than other methods with improved accuracy.

\subsection{Ablation Test for Shift-and-Scale Operation}
To evaluate the effect by shifting and scaling the last layer according to the sub-block type resolved value distribution (i.e., the mean values for scalars and standard deviation for both scalars and vectors), we compared the average values of the MAE matrix using DeepH-E3, the pristine MACE-H, and the MACE-H with shift-and-scale operation. In Supplementary Fig. \ref{fig: SS comparison}a, the shift-and-scale operation of MACE-H reduced the average MAE error from 0.32 meV of the pristine one to 0.27 meV. Due to the small magnitude of the SOC-involved matrix sub-blocks between different spin channels, the corresponding minimum value of the error matrix decreases by 3 magnitudes. A similar trend can also be observed for bulk Au in Supplementary Fig. \ref{fig: SS comparison}b.

\begin{figure}[htbp!]
    \centering\includegraphics[width=0.8\textwidth,height=0.8\textheight,keepaspectratio] {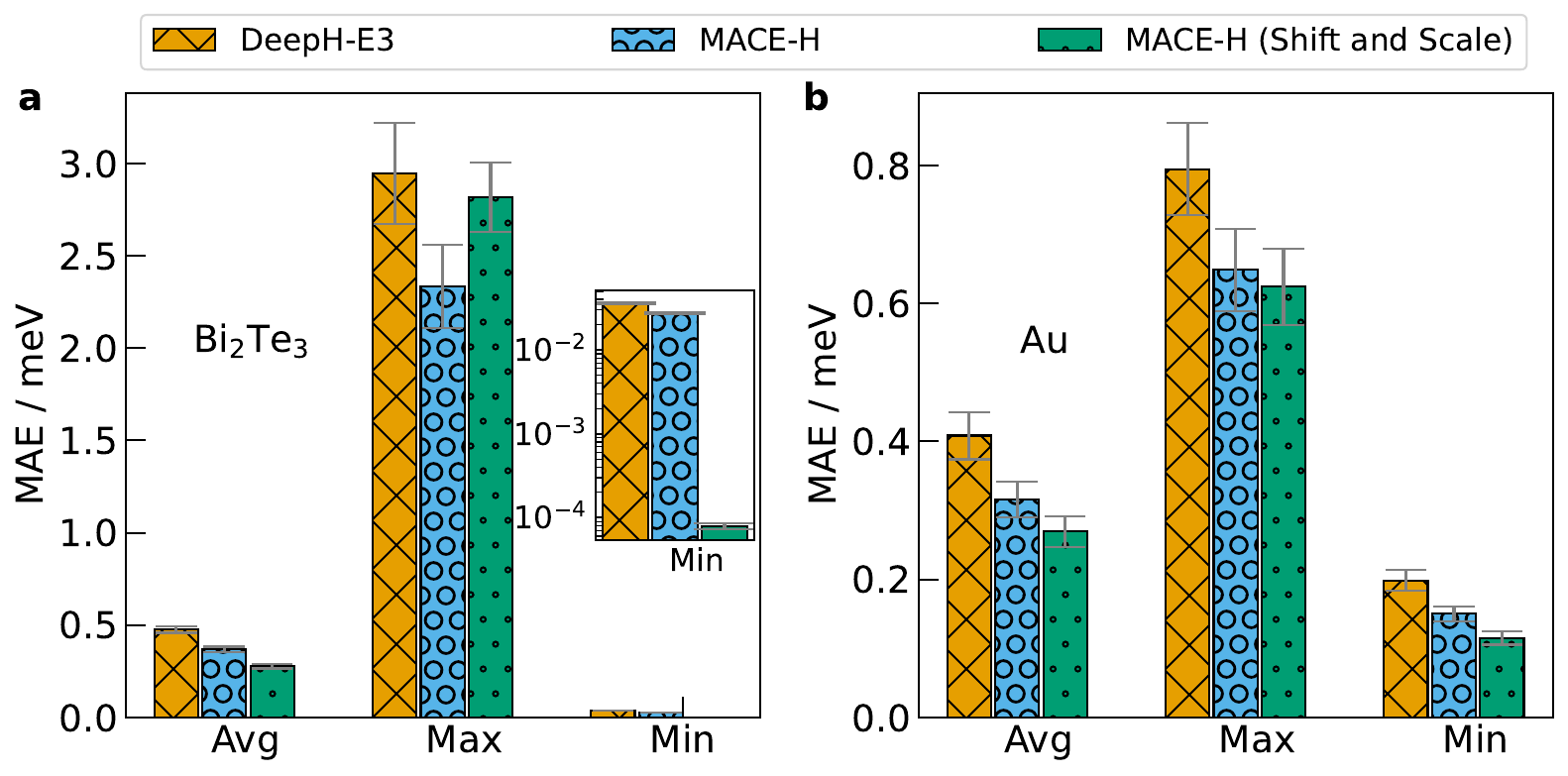}
    \caption{\textbf{Average values of the Hamiltonian error matrices.} Data shown for (\textbf{a}) Bi\textsubscript{2}Te\textsubscript{3} bilayer and (\textbf{b}) bulk Au using DeepH-E3, the pristine MACE-H, and the MACE-H with the shift-and-scale operation. The confidence interval of a standard deviation $\sigma$ is taken across the error matrix metric of individual samples. The inset in \textbf{a} is the magnified view of the Min MAE for Bi\textsubscript{2}Te\textsubscript{3}}
    \label{fig: SS comparison}
\end{figure}

\section{The correlation between hermiticity error and prediction error}\label{sec:hermiticity_error}
To further justify the correlation between the hermiticity error and the prediction error, which can be useful for future active learning methods, we first compared the median numbers of the prediction error of the matrix blocks with different quantiles of the hermiticity error in Supplementary Fig.~\ref{fig:hermiticity_quantile}. For the matrix blocks that have higher hermiticity errors in the ranking (lower quantile value), the corresponding prediction errors (measured by the median number of the corresponding matrix blocks) are usually higher than those with lower hermiticity errors.

Moreover, since the new label selection during active learning in practice is usually based on the entire geometry rather than single matrix block components, we compared the conformation-wise prediction error with the hermiticity error for models trained with varying training data sizes in Supplementary Fig.~\ref{fig:sample_hermiticity}. The configuration-wise hermiticity errors are calculated as the mean of the block-wise errors within each configuration. For models trained with different numbers of conformations, the conformation-wise prediction errors are positively correlated with the hermiticity error, demonstrating the universality of this observation. This offered the possibility for an efficient measurement for active learning.

\begin{figure}[htbp!]
    \centering
    \includegraphics[width=0.6\linewidth]{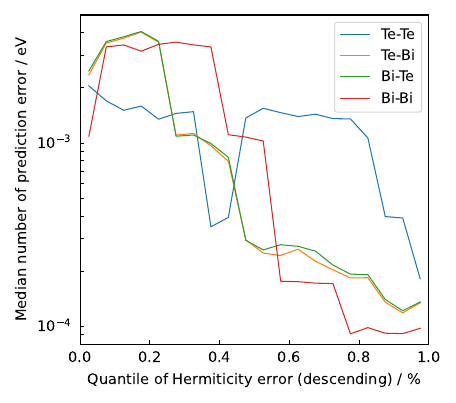}
    \caption{\textbf{The comparison of the median number of prediction errors for matrix blocks with corresponding quantile of hermiticity error in the twisted Bi\textsubscript{2}Te\textsubscript{3} configuration resolved for different element pairs.} The quantile values here are in descending order, i.e., lower quantiles indicate higher hermiticity error.}
    \label{fig:hermiticity_quantile}
\end{figure}

\begin{figure}[htbp!]
    \centering
    \includegraphics[width=\linewidth]{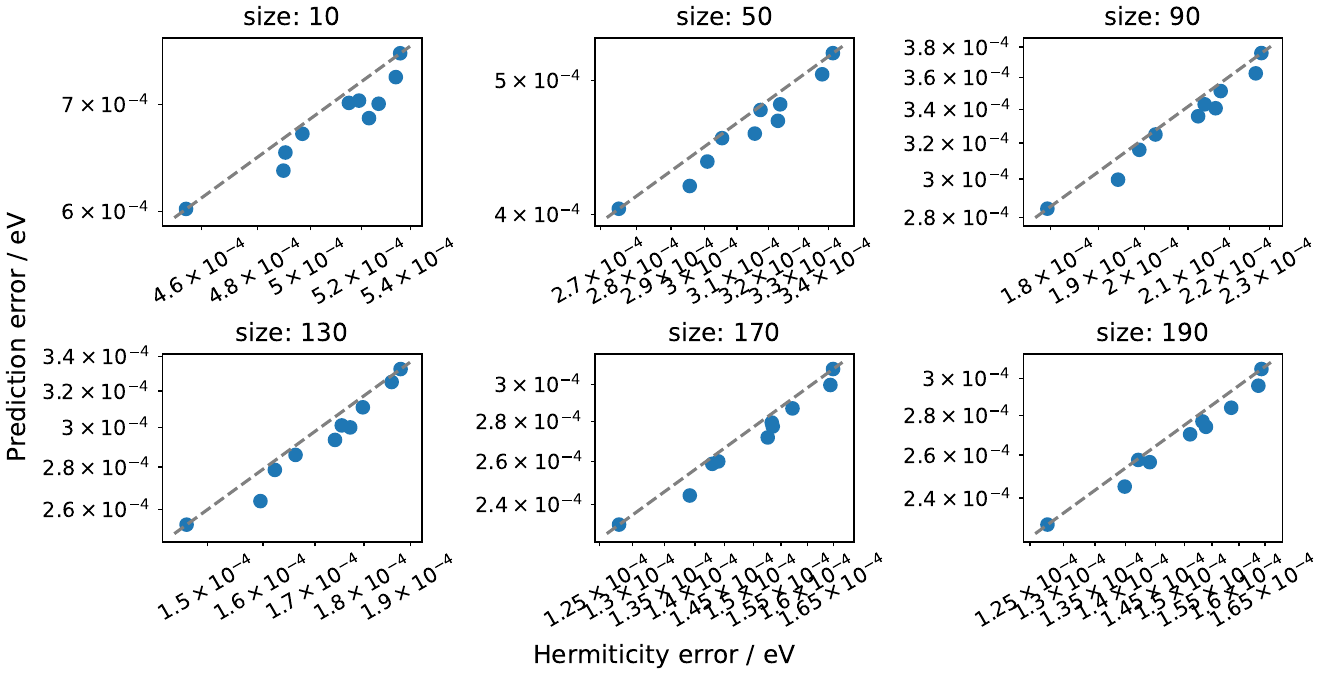}
    \caption{\textbf{The configuration-wise prediction error versus hermiticity error of held-out configurations for the gold models trained with varying training data sizes (as denoted by "size" in the subplots).} The configuration-wise hermiticity errors are calculated as the mean of the block-wise errors within each configuration.}
    \label{fig:sample_hermiticity}
\end{figure}

\section{Core Projection and Sparsification}\label{sec:si_coreproj}

\subsection{Reciprocal-to-Real Transformation Convergence}\label{sec:si_coreproj_reciptoreal}

It was investigated how many k-points are required to accurately perform the inverse Fourier transform from reciprocal to real space for core-projected Hamiltonian and overlap matrices, as shown in Supplementary Fig.~\ref{fig:si_coreproj_reciptoreal}. The convergence study was performed for a single configuration in the training set called 'run\_Aims\_duy8hd\_p', which can be found on NOMAD. Based on these results, it was decided to perform core projection for the whole gold dataset using the 6x6x6 k-point grid. Note that the reciprocal cell for each configuration in the gold dataset is the same and contains reciprocal-space vectors of equal length.

\begin{figure}[htbp!]
    \centering
    \includegraphics[width=0.6\linewidth]{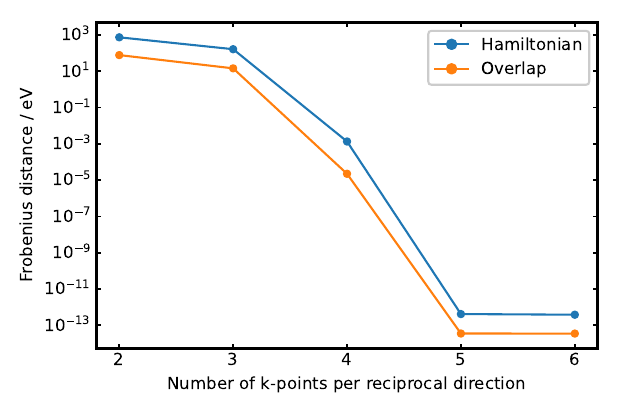}
    \caption{\textbf{Convergence of reciprocal-to-real transformation (inverse Fourier transform) as a function of k-point grid for core-projected Hamiltonian and overlap matrices for one of the configurations in the training set.} The Frobenius distance corresponds to $\Vert \bm{H}_{n_{k}} - \bm{H}_{n_{k,\mathrm{ref}}}\Vert_{\mathcal{F}}$, where $n_{k,\mathrm{ref}}$ is the reference number of k-points per reciprocal dimension, which was set to 7, corresponding to a $7\times7\times7$ k-point grid.}
    \label{fig:si_coreproj_reciptoreal}
\end{figure}

\subsection{Eigenvalue Accuracy:}\label{sec:si_eigenacc}

Core states from Hamiltonian and overlap matrices in the gold dataset were projected out as discussed in the main text and Sec.~\ref{sec:si_coreproj_reciptoreal}. Furthermore, to decrease the computational cost during training and inference, the off-site matrix blocks corresponding to interactions spanning over 10~Å were set to zero. This was motivated by the fact that absolute Hamiltonian and overlap matrix elements for edges beyond 10~Å become too small to significantly affect Hamiltonian eigenvalues as quantitatively discussed below.

It was examined how accurate the resulting eigenvalues from processed (core-projected and sparsified) matrices are compared to eigenvalues obtained from original full-basis matrices. This was investigated for a single gold configuration in the training set ('run\_Aims\_duy8hd\_p'). It was found that the eigenvalue and electronic entropy errors at 1000 K between the full and processed matrices are equal to $1.41 \times 10^{-4}$ eV and $6.96 \times 10^{-12}$ eV K$^{-1}$ Å$^{-3}$, respectively. These values are much smaller than the model errors shown in Supplementary Fig. 4 in the main text. The valence electronic band structures using full and processed matrices are shown in Supplementary Fig.~\ref{fig:si_eigenacc}

\begin{figure}[htbp!]
    \centering
    \includegraphics[width=0.72\linewidth]{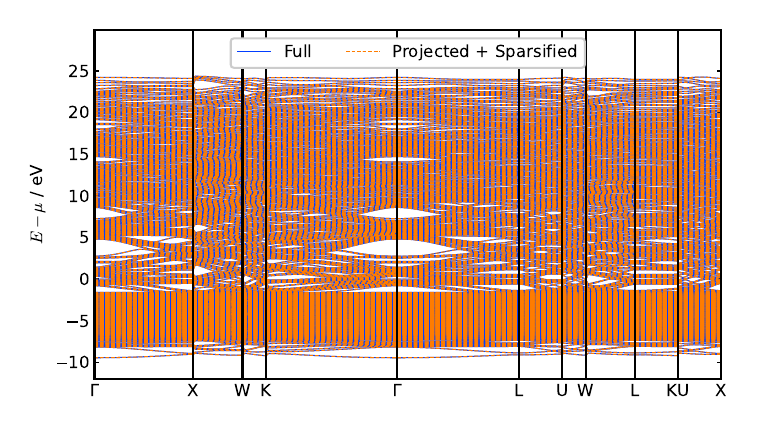}
    \caption{\textbf{Valence electronic band structure using full and processed (core-projected + sparsified) matrices for one of the configurations in the training set.} The bands from full and processed matrices are shifted with respect to their chemical potentials $\mu$ obtained with Fermi-Dirac smearing at 1000~K.}
    \label{fig:si_eigenacc}
\end{figure}

\section{Additional data}\label{sec:benchmark}

All data below (except where stated otherwise) employs MACE-H models without shift-and-scale operation.

\begin{table}[htbp!]
\centering
\caption{Hyperparameters of the models for different datasets in Fig. 4. The $e_{\mathrm{N}}$, $l_r$, $N_B$, $N_{\mathrm{layers}}$, $\nu$, $L_{\mathrm{MACE}}$, $L_{\mathrm{node}}$, $L_{\mathrm{hidden}}$, and $L_{\mathrm{edge}}$ stand for the node-update radial basis (B for Bessel and G for Gaussian), learning rate, batch size, number of layers, correlation order, the largest azimuthal number of the intermediate MACE block irreps, the intermediate node-wise block irreps, the many-body expansion hidden states irreps, and the intermediate edge-wise block irreps.}
\resizebox{\textwidth}{!}{
\begin{tabular}{c c c c c c c c c c c c c c c} 
\toprule
\multirow{2}{*}{Dataset} & \multicolumn{8}{c}{MACE-H} & \multicolumn{6}{c}{DeepH-E3} \\ 
\cmidrule(lr){2-9} \cmidrule(lr){10-15}
& $e_{\mathrm{N}}$ & $l_r$ & $N_B$ & $N_{\mathrm{layers}}$ & $\nu$ & $L_{\mathrm{MACE}}$ & $L_{\mathrm{hidden}}$ & $L_{\mathrm{edge}}$ & $e_{\mathrm{N}}$ & $l_r$ & $N_B$ & $N_{\mathrm{layers}}$ & $L_{\mathrm{node}}$ & $L_{\mathrm{edge}}$ \\
\midrule
monolayer graphene (noSOC) & G & 0.006 & 1 & 3 & 3 & 5 & 5 & 5 & G & 0.003 & 1 & 3 & 5 & 5 \\
monolayer MoS\textsubscript{2} (noSOC) & G & 0.006 & 1 & 3 & 3 & 5 & 5 & 5 & G & 0.005 & 1 & 3 & 5 & 5 \\
bilayer graphene (noSOC) & G &  0.003 & 1 & 3 & 3 & 5 & 5 & 5 & G & 0.003 & 1 & 3 & 5 & 5 \\
bilayer Bismuthene (SOC) & G &  0.01 & 1 & 3 & 3 & 5 & 5 &  5 & G & 0.005 & 1 & 3 & 5 & 5 \\
bilayer Bi\textsubscript{2}Se\textsubscript{3} (SOC) & G & 0.01 & 1 & 3 & 3 & 5 & 5 & 5 & G & 0.005 & 1 &  3 & 5 & 5 \\
bilayer Bi\textsubscript{2}Te\textsubscript{3} (SOC) & G & 0.008 & 2 & 3 & 3 & 5 & 5 & 5 & G & 0.004 & 2 & 3 & 5 & 5 \\
bilayer Bi\textsubscript{2}Te\textsubscript{3} (noSOC) & G & 0.008 & 2 & 3 & 3 &  5 & 2 & 4 & G & 0.004 & 2 & 3 & 5 & 5 \\
bulk Au (noSOC) & B & 0.008 & 2 & 2 & 2 & 5 & 2 & 4 & G & 0.008 & 2 & 2 & 4 & 4 \\
\bottomrule
\end{tabular}
}
\label{table: hyperparameters}
\end{table}

\begin{table}[htbp!]
\centering
\caption{MAE (in meV) for Hamiltonian matrix prediction on 2D monolayers without SOC using DeepH, DeepH-E3 and our MACE-H. The basis sets used for both graphene and MoS\textsubscript{2} are up to \textit{d}-orbital.}
\resizebox{0.65\textwidth}{!}{
\begin{tabular}{c c c c c c} 
\toprule
 \multirow{2}{*}{model} & Graphene (noSOC) & \multicolumn{4}{c}{MoS\textsubscript{2} (noSOC)} \\ 
\cmidrule(lr){2-2} \cmidrule(lr){3-6}
& C-C & Mo-Mo & Mo-S & S-S & overall \\
\midrule
 DeepH & 2.1 & 1.3 & 0.9 & 0.7 & 0.95 \\ 
 DeepH-E3 & 0.27 & 0.51 & 0.44 & 0.36 & 0.45 \\
 MACE-H (ours) & \textbf{0.21} & \textbf{0.42} & \textbf{0.39} & \textbf{0.32} & \textbf{0.39} \\
\bottomrule
\end{tabular}
}
\label{table:benchmark_monolayer}
\end{table}

\begin{table}[htbp!]
\centering
\caption{MAE (in meV) for Hamiltonian matrix prediction on 2D bilayers using DeepH, DeepH-E3 and our MACE-H. The basis sets are up to \textit{d}-orbital, while the graphene and Bi\textsubscript{2}Te\textsubscript{3} dataset doesn't involve SOC, the Bismuthene, Bi\textsubscript{2}Te\textsubscript{3}, Bi\textsubscript{2}Se\textsubscript{3} involve SOC.}
\resizebox{\textwidth}{!}{
\begin{tabular}{c c c c c c c c c c c} 
\toprule
 \multirow{2}{*}{model} & \multicolumn{2}{c}{Graphene (noSOC)} & \multicolumn{2}{c}{Bi\textsubscript{2}Te\textsubscript{3} (noSOC)} & \multicolumn{2}{c}{Bismuthene (SOC)} & \multicolumn{2}{c}{Bi\textsubscript{2}Te\textsubscript{3} (SOC)} & \multicolumn{2}{c}{Bi\textsubscript{2}Se\textsubscript{3} (SOC)} \\
\cmidrule(lr){2-3} \cmidrule(lr){4-5} \cmidrule(lr){6-7} \cmidrule(lr){8-9}
& shifted & twisted & shifted & twisted & shifted & twisted & shifted & twisted \\
\midrule
 DeepH & 1.9 & 0.62 & / & / & / & / & / & / & / & / \\ 
 DeepH-E3 & 0.40 & 0.28 & 0.86 & \textbf{1.56} & 0.26 & \textbf{0.40} & 0.48 & \textbf{0.70} & 0.40 & 0.35 \\
 MACE-H (ours) & \textbf{0.31} & \textbf{0.24} & \textbf{0.58} & 2.50 & \textbf{0.23} & 0.45 & \textbf{0.37} & 1.25 & 0.34 & 0.35 \\
\bottomrule
\end{tabular}
}
\label{table:benchmark_bilayer}
\end{table}

\begin{table}[htbp!]
\centering
\caption{MAE (in meV) on bilayer Bi\textsubscript{2}Te\textsubscript{3} with SOC using different settings of radial basis and many-body expansion and its relation to the locality. The values in the parentheses represent the maximum value of the MAEs among different element pair-resolved orbital pairs, indicating the model performance for the most challenging orbital pair, which is the same as that in the following Tables S5 and S6. The B and G in parentheses stand for Bessel and Gaussian radial basis set, respectively.}
\resizebox{\textwidth}{!}{
\begin{tabular}{c c c c c c c c c} 
\toprule
 \multirow{2}{*}{model} & \multicolumn{4}{c}{shifted Bi\textsubscript{2}Te\textsubscript{3} (SOC)} & \multicolumn{4}{c}{twisted Bi\textsubscript{2}Te\textsubscript{3} (SOC)} \\ 
\cmidrule(lr){2-5} \cmidrule(lr){6-9}
& Te-Te & Te-Bi & Bi-Bi & overall & Te-Te & Te-Bi & Bi-Bi & overall \\
\midrule
 DeepH-E3 & 0.63 (3.23) & 0.48 (2.34) & 0.41 (2.08) & 0.50 (3.23) & 0.90 (11.84) & 0.59 (4.74) & 0.66 (7.11) & 0.68 (11.84)\\ 
 MACE-H (B, \textit{v}=3, \textit{L}=2) & 0.51 (3.08) & 0.37 (1.84) & 0.34 (1.75) & 0.40 (3.80) & 1.26 (17.91) & 1.10 (12.10) & 1.51 (23.83) & 1.24 (23.83)\\
 MACE-H (G, \textit{v}=3, \textit{L}=5) & 0.46 (2.33) & 0.35 (1.66) & 0.30 (1.62) & 0.37 (2.33) & 1.26 (17.91) & 1.10 (12.10) & 1.51 (23.83) & 1.24 (23.83) \\
 MACE-H (G, \textit{v}=3, \textit{L}=2) & 0.50 (2.49) & 0.38 (1.80) & 0.34 (1.66) & 0.40 (2.49) & 1.16 (18.11) & 1.06 (9.19) & 1.48 (20.93) & 1.19 (20.93) \\
 MACE-H (G, \textit{v}=1, \textit{L}=1) & 0.57 (2.87) & 0.43 (2.04) & 0.38 (1.89) & 0.45 (2.87) & 1.03 (12.70) & 0.82 (6.42) & 0.98 (12.46) & 0.91 (12.70) \\
\bottomrule
\end{tabular}
}
\label{table:benchmark_monolayer}
\end{table}

\begin{table}[htbp!]
\centering
\caption{MAE (in meV) on bilayer Bi\textsubscript{2}Se\textsubscript{3} with SOC using different settings of radial basis and many body expansion and its relation to the locality. The values in the parentheses show the maximum error of the Hamiltonian matrix elements. The B and G in parentheses stand for Bessel and Gaussian radial basis set, respectively.}
\resizebox{\textwidth}{!}{
\begin{tabular}{c c c c c c c c c} 
\toprule
 \multirow{2}{*}{model} & \multicolumn{4}{c}{shifted Bi\textsubscript{2}Se\textsubscript{3} (SOC)} & \multicolumn{4}{c}{twisted Bi\textsubscript{2}Se\textsubscript{3} (SOC)} \\ 
\cmidrule(lr){2-5} \cmidrule(lr){6-9}
& Se-Se & Se-Bi & Bi-Bi & overall & Se-Se & Se-Bi & Bi-Bi & overall \\
\midrule
 DeepH-E3 & 0.39 (1.45) & 0.45 (1.72) & 0.32 (1.97) & 0.40 (1.97) & 0.37 (1.58) & 0.38 (1.71) & 0.30 (2.29) & 0.35 (2.29)\\
 MACE-H (B, \textit{v}=3, \textit{L}=2) & 0.35 (1.33) & 0.39 (1.57) & 0.28 (1.69) & 0.35 (1.69) & 0.38 (1.95) & 0.39 (1.84) & 0.28 (2.41) & 0.35 (2.41) \\
 MACE-H (G, \textit{v}=3, \textit{L}=2) & 0.33 (1.20) & 0.37 (1.49) & 0.29 (1.64) & 0.34 (1.64) & 0.42 (2.29) & 0.52 (2.89) & 0.47 (3.02) & 0.48 (3.02) \\
 MACE-H (G, \textit{v}=3, \textit{L}=5) & 0.32 (1.25) & 0.37 (1.40) & 0.28 (1.69) & 0.33 (1.69) &  0.38 (2.37) & 0.43 (2.41) & 0.46 (3.38) & 0.43 (3.38) \\
 MACE-H (G, \textit{v}=1, \textit{L}=1) & 0.35 (1.39) & 0.39 (1.57) & 0.29 (1.54) & 0.35 (1.59) & 0.40 (2.07) & 0.46 (2.33) & 0.46 (4.04) &  0.45 (4.04) \\
\bottomrule
\end{tabular}
}
\label{table:benchmark_monolayer}
\end{table}

\begin{table}[htbp!]
\centering
\caption{The MAE (in meV) on Au data using DeepH-E3, MACE-H with/without shift-and-scale operation. The values in the parentheses show the maximum error of the Hamiltonian matrix elements.}
\footnotesize
\begin{tabular}{c c c c} 
\toprule
     & DeepH-E3 & MACE-H & MACE-H (Shift \& Scale) \\
\midrule
Au & 0.41 (0.80) & 0.32 (0.65) & 0.27 (0.62) \\
\bottomrule
\end{tabular}
\label{table: Au datasets}
\end{table}

\begin{table}[htbp!]
\centering
\caption{MAE (in meV) on bilayer graphene without SOC and Bismuthene with SOC using different settings of radial basis and many body expansion and its relation to the locality. Values in parentheses show the maximum error of the Hamiltonian matrix elements. The B and G in parentheses stand for Bessel and Gaussian radial basis set, respectively.}
\normalsize
\resizebox{0.75\textwidth}{!}{
\begin{tabular}{c c c c c} 
\toprule
 \multirow{2}{*}{model} & \multicolumn{2}{c}{graphene (noSOC)} & \multicolumn{2}{c}{Bismuthene (SOC)} \\ 
\cmidrule(lr){2-3} \cmidrule(lr){4-5}
& shifted & twisted & shifted & twisted \\
\midrule
 DeepH-E3 & 0.40 (0.73) & 0.28 (1.04) & 0.26 (1.55) & 0.40 (3.94)\\ 
 MACE-H (B, \textit{v}=3, \textit{L}=5) & 0.29 (0.52) & 0.24 (0.78) & 0.25 (1.4) & 0.52 (6.29)\\
 MACE-H (G, \textit{v}=3, \textit{L}=5) & 0.31 (0.51) & 0.24 (0.80) & 0.23 (1.27) & 0.45 (4.79) \\
 MACE-H (G, \textit{v}=1, \textit{L}=1) & 0.40 (0.75) & 0.24 (0.59) & 0.24 (1.50) & 0.37 (3.39) \\
\bottomrule
\end{tabular}
}
\label{table:benchmark_monolayer}
\end{table}

\begin{table}[htbp!]
\centering
\caption{Summary of employed datasets.}
\resizebox{\textwidth}{!}{
\begin{tabular}{c c c c c c c} 
\toprule
Dataset & system dimension & DFT package & number of layers & SOC & data size (train/val/test) & system size (atoms) \\
\midrule
graphene & 2D & OpenMx & monolayer & noSOC & 450 (0.6/0.2/0.2) & 72 \\
MoS\textsubscript{2} & 2D & OpenMx & monolayer & noSOC & 500 (0.6/0.2/0.2) & 75 \\
graphene & 2D & OpenMx & bilayer & noSOC & 300 (0.6/0.2/0.2) & 64 \\
bismuthene & 2D & OpenMx & bilayer & SOC & 576 (0.4/0.2/0.2) & 36 \\
Bi\textsubscript{2}Se\textsubscript{3} & 2D & OpenMx & bilayer & SOC & 576 (0.4/0.2/0.2) & 90 \\
Bi\textsubscript{2}Te\textsubscript{3} & 2D & OpenMx & bilayer & noSOC & 256 (0.8/0.15/0.05) & 90 \\
Bi\textsubscript{2}Te\textsubscript{3} & 2D & OpenMx & bilayer & SOC & 256 (0.8/0.15/0.05) & 90 \\
Au & Bulk & FHI-aims & / & noSOC & 200 (0.76/0.19/0.05) & 64 \\
\bottomrule
\end{tabular}
}
\label{table: datasets summary}
\end{table}

\begin{table}[htbp!]
\centering
\caption{Radial cutoffs of numeric atom-centered orbitals (NAOs) in the employed datasets. The values can be used to infer the sparsity of the Hamiltonian and, therefore, the resulting connectivity graph.}
\footnotesize
\begin{tabular}{c c c c c c c c} 
\toprule
& C & Mo & S & Bi & Se & Te & Au\\
\midrule
Radius & 6.0 a$_0$ & 7.0 a$_0$ & 7.0 a$_0$ & 8.0 a$_0$ & 7.0 a$_0$ & 7.0 a$_0$ & 5.0 \AA{}\\
\bottomrule
\end{tabular}
\label{table: nao cutoffs}
\end{table}